\documentclass[a4paper,UKenglish]{lipics-v2018}

\usepackage{microtype}

\usepackage{amsfonts}
\usepackage{caption}
\usepackage{xspace}
\usepackage{mathtools}
\usepackage{mathpartir}
\usepackage{ifpdf}
\usepackage{graphicx}
\usepackage{stmaryrd}
\usepackage{wrapfig}
\usepackage{textcomp}
\usepackage{color}
\usepackage{url}
\usepackage[all,cmtip]{xy}
\usepackage{algorithm}
\usepackage{algpseudocode}
\usepackage{empheq}

\algnewcommand\algorithmicswitch{\textbf{match}}
\algnewcommand\algorithmiccase{\textbf{case}}
\algnewcommand\algorithmicwith{\textbf{with}}

\algdef{SE}[MATCH]{Match}{EndMatch}[1]{\algorithmicswitch\ #1\ \algorithmicwith}{\algorithmicend\ \algorithmicswitch}%
\algdef{SE}[CASE]{Case}{EndCase}[1]{\algorithmiccase\ #1}{\algorithmicend\ \algorithmiccase}%
\algtext*{EndMatch}%
\algtext*{EndCase}%

\newcommand{\ALT}{~\mid~}
\newcommand{\RULE}[2]{\frac{\begin{array}{c}#1\end{array}}
                           {\begin{array}{c}#2\end{array}}}
\newcommand{\rulelabel}[1]{\textrm{\sc {#1}}}
\newcommand{\C}[1]{\texttt{#1}}
\newcommand{\F}[1]{\mathsf{#1}}
\newcommand{\M}[1]{\mathcal{#1}}

\title{Automated Detection of Serializability Violations under Weak Consistency (Extended Version)}

\titlerunning{Automated Detection of Serializability Violations under Weak Consistency}

\author{Kartik Nagar and Suresh Jagannathan}{Purdue University, USA}{nagark@purdue.edu, suresh@cs.purdue.edu}{}{}

\authorrunning{K.Nagar and S.Jagannathan}

\Copyright{Kartik Nagar and Suresh Jagannathan}

\subjclass{Theory of computation $\rightarrow$ Automated reasoning}


\category{}

\relatedversion{}

\supplement{}

\funding{}

\EventEditors{Sven Schewe and Lijun Zhang}
\EventNoEds{2}
\EventLongTitle{29th International Conference on Concurrency Theory (CONCUR 2018)}
\EventShortTitle{CONCUR 2018}
\EventAcronym{CONCUR}
\EventYear{2018}
\EventDate{September 4--7, 2018}
\EventLocation{Beijing, China}
\EventLogo{}
\SeriesVolume{118}
\ArticleNo{41} 
\nolinenumbers 
\hideLIPIcs  

\begin{document}

\maketitle

\begin{abstract}

  While a number of weak consistency mechanisms have been developed in
  recent years to improve performance and ensure availability in
  distributed, replicated systems, ensuring correctness of
  transactional applications running on top of such systems remains a
  difficult and important problem. Serializability is a
  well-understood correctness criterion for transactional programs;
  understanding whether applications are serializable when executed in
  a weakly-consistent environment, however remains a challenging
  exercise.  In this work, we combine the dependency graph-based characterization of serializability and the framework of
  abstract executions to develop a fully automated approach for
  statically finding bounded serializability violations under
  \emph{any} weak consistency model. We reduce the problem of
  serializability to satisfiability of a formula in First-Order Logic,
  which allows us to harness the power of existing SMT solvers. We
  provide rules to automatically construct the FOL encoding from
  programs written in SQL (allowing loops and conditionals) and the
  consistency specification written as a formula in FOL. In addition
  to detecting bounded serializability violations, we also provide two
  orthogonal schemes to reason about unbounded executions by providing
  sufficient conditions (in the form of FOL formulae) whose
  satisfiability would imply the absence of anomalies in any arbitrary
  execution. We have applied the proposed technique on TPC-C, a real
  world database program with complex application logic, and were able
  to discover anomalies under Parallel Snapshot Isolation, and verify
  serializability for unbounded executions under Snapshot Isolation,
  two consistency mechanisms substantially weaker than serializability.
\end{abstract}

\section{Introduction}

We consider the problem of detecting serializability violations of
transactional programs executing in a weakly-consistent replicated
distributed database.  An execution of such programs is said to be
\emph{serializable} if it is equivalent to a serial, sequential
execution of the transactions that comprise the program.  Ensuring
that all executions of such programs are serializable greatly
simplifies reasoning about program correctness by reducing the
complexity of understanding concurrent executions to the problem of
understanding sequential ones.  Unfortunately, \emph{enforcing}
serializability using runtime synchronization mechanisms is
problematic in geo-replicated distributed systems without sacrificing
availability (low-latency)~\cite{GI02}.  To reap the correctness
benefits of serializability with the performance and scalability
benefits of high-availability, we study the conditions under which
transactional programs can be statically identified to always yield a
serializable execution \emph{without} the need for global
synchronization.  The challenge to realizing this goal stems from the
complexity in reasoning about replicated state in which not all
replicas share the same view of the data they hold.

To address this challenge, we present a fully automated static
analysis that precisely encodes salient dependencies in the program as
abstract executions defined in terms of an axiomatic specification of
a particular weak consistency model (\S 4). The analysis then leverages a theorem
prover to systematically search for the presence or absence of cycles
in these executions consistent with these dependencies; the presence
of a cycle indicates a serializability violation (\S 5.1).  Notably, our
approach can be applied to any weak consistency model whose
specification can be expressed in first-order logic, a class that
subsumes all realistic data stores we are aware of.  More
specifically, our approach constructs a dependency graph \cite{AD00} from the
input program containing a cycle and then asks whether there exists a
valid execution under the given consistency specification that can
result in this graph. To do this, we automatically extract from the
transactional program conditions under which dependencies can occur,
and relate the dependencies to artifacts in an event-based model to
find whether there exists a valid abstract execution corresponding to
the dependency graph. These dependencies are encoded in a first-order
logic formula that is satisfiable only if there exists an execution that
violates serializability.

Given a transactional program written in SQL, we discover
serializability violations of bounded length under the given weak
consistency model (with the bound limiting the number of concurrent
transaction instances that are considered). We output the actual
anomaly including the transactions involved and their inputs. This
output can then be used to strengthen the consistency of the
transactions involved in the anomaly (or even modifying the
transactions themselves).  Since the approach is parametric on a
consistency policy, it can also be used to determine the weakest
consistency policy for which the program is serializable.  Consistent
with other bounded verification techniques used to detect bugs in
e.g., concurrent programs~\cite{MU08}, we posit that most
serializability violations will manifest using a small number of
transaction instances.  

We provide two orthogonal schemes to reason about arbitrarily long
executions with an unbounded number of transaction instances (\S 5.2, \S 5.3). The first
scheme formalizes the argument that it is enough to check
serializability violations in bounded executions, by proving that
longer violations beyond that bound would induce violations within the
bound.  The second scheme applies an inductive argument to check the
absence of anomalies in arbitrarily long executions.  Our approach is
sound, but not complete - while all discovered anomalies are justified
by counterexamples offered by the theorem prover, we cannot rule out the
possibility of serializability violations appearing in unbounded
executions that are not identified by these two schemes.

As serious case studies to assess the applicability of our approach,
we have applied our technique on TPC-C, a real-world transactional
program and a Courseware application  (\S 6). In both cases, we were
able to detect multiple serializability violations under Eventual Consistency and a weaker
variant of Snapshot Isolation (SI) called Parallel Snapshot
Isolation~\cite{SO11}, and verified that these anamolies did not
occur when using SI for unbounded executions. We now present an overview of our approach using a simple example.

\section{Overview}
\begin{figure}
\begin{lstlisting}
withdraw (ID, Amount)
    SELECT Balance AS bal WHERE AccID=ID
    IF bal > Amount
        UPDATE SET Balance=bal-Amount WHERE AccID=ID 
\end{lstlisting}
\caption{Example Application}
\label{Fig:1}
\vspace{-10pt}
\end{figure}
In this section, we show how our approach discovers serializability violations, and how the output of
our analysis can be used to repair violations using selective
synchronization.  Consider a simple banking application which
maintains the balance of multiple accounts in a table \C{Account}
which is indexed using the primary key \C{AccID} and contains the
field \C{Balance}. Consider a \C{withdraw} operation (shown in Fig.
\ref{Fig:1}) written in a SQL-style language, which takes \C{ID} and
\C{Amount} as input, and deducts the amount from the account with
account number \C{ID} if the balance is sufficient.  Suppose the
application is deployed in a distributed, replicated environment which
allows concurrent invocations of the \C{withdraw} operation at
potentially different replicas, with the only guarantee provided being
eventual consistency - eventually, all replicas will witness all
updates to the \C{Balance} field.  Under eventual consistency, the
application is clearly not serializable, since concurrent withdraws
operations to the same account--whose total withdrawn amount exceeds
the balance of the account--could both succeed, which is not possible
in a serializable execution.

A convenient way to express executions in such an environment is to
use an axiomatic event-based representation. In this framework, an
abstract execution \cite{CE15} is expressed as the tuple $(T, \F{vis}, \F{ar})$,
where $T$ is the set of transaction invocations, $\F{vis} \subseteq T \times T$ is a visibility
relation such that if $t \xrightarrow{\F{vis}} t'$ then updates of $t$ are
visible to $t'$, and $\F{ar} \subseteq T \times T$ is an arbitration relation which totally
orders all writes to the same location and ensures eventual
consistency \cite{BU12}. For example, if $t_1=$ \C{withdraw}(1,50), $t_2 =$
\C{withdraw}(1,60), then $E = (\{t_1, t_2\}, \{\}, \{(t_1, t_2)\})$ is
an abstract execution which is not serializable, because the final value
of \C{Balance} in the account number 1 will only reflect the
\C{withdraw} operation $t_2$ (assuming an initial \C{Balance} of 100
in \C{AccID} 1), since there is no visibility constraint enforced
between the two operations. This is an example of a \emph{lost update} \cite{BE16}
anomaly. Our goal is to automatically construct such anomalous
executions.

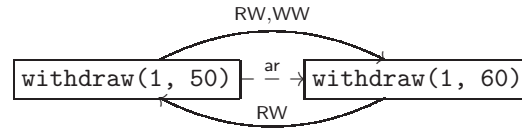
\begin{wrapfigure}{r}{0.5 \textwidth}
\vspace{-20pt}
$
\xymatrix{*+[F]{\C{withdraw(1, 50)}} \ar@{-->}[r]^{\F{ar}} \ar@/^1.5pc/[r]^{\F{RW}, \F{WW}}   &   *+[F]{\C{withdraw(1, 60)}} \ar@/^1.5pc/[l]_{\F{RW}} }
$
\caption{Abstract Execution $E$ and its Dependency Graph}
\label{Fig:2}
\vspace{-20pt}
\end{wrapfigure}

A useful technique to detect serializability violations is to build
dependency graphs from abstract executions, and then search for cycles
in the dependency graph. The nodes of the dependency graph are
invocations, and edges indicate dependencies between them. There are
three type of dependencies relevant to serializability detection: $t_1
\xrightarrow{\F{WR}} t_2$ is a read dependency, which means that $t_2$
reads a value written by $t_1$, $t_1 \xrightarrow{\F{WW}} t_2$ is a
write dependency, which means that both $t_1$ and $t_2$ write to the
same location, with the write of $t_2$ arbitrated after $t_1$, and
$t_1 \xrightarrow{\F{RW}} t_2$ is an anti-dependency, which means that
$t_1$ does \emph{not} read a value written by $t_2$ but instead reads an
older version. For example, the dependency graph of the anomalous
execution $E$ described above is shown in Fig. \ref{Fig:2}.

In our approach, we start with a dependency graph containing a cycle,
and then ask whether an execution corresponding to the dependency
graph is possible. From the transaction code, we automatically extract
the conditions under which a dependency edge can manifest between
invocations of the transactions. In our running example, a dependency
edge (of any type) between two \C{withdraw} invocations can only manifest if they
are called with the same account ID. Further, we link the dependency
edges with the relations $\F{vis}$ and $\F{ar}$ of the corresponding
abstract execution. For example, $t_1 \xrightarrow{RW} t_2 \Rightarrow
\neg (t_2 \xrightarrow{vis} t_1)$, because otherwise, $t_1$ would read
the value written by $t_2$. This is useful because different
consistency schemes can be axiomatically expressed by placing
constraints on $\F{vis}$ and $\F{ar}$ relations.

In order to prevent the anomalous execution in our running example, we
can use Parallel Snapshot Isolation \cite{SO11} which ensures that if two
invocations write to the same location, then they cannot be
concurrent. While PSI is implemented using a complex, distributed
protocol, in our abstract framework, it can be simply expressed using
the following constraint : $\forall t,t'.\ t \xrightarrow{\F{WW}} t'
\implies t \xrightarrow{\F{vis}} t'$. Now, the anomalous execution $E$ is
not possible, because $t_1 \xrightarrow{\F{WW}} t_2 \Rightarrow t_1
\xrightarrow{\F{vis}} t_2$, which contradicts $t_2
\xrightarrow{\F{RW}} t_1$.

To summarize, the following is the relevant portion of formulae
that we generate for the above application under PSI:
\begin{gather}
\forall t,t'.\ t \xrightarrow{\F{RW}} t' \Rightarrow (\exists r.\ \C{AccID}(r) = \C{ID}(t) \wedge \C{AccID}(r) = \C{ID}(t') \wedge \C{bal}(t') > \C{Amount}(t'))\\
\begin{split}
\forall t,t',r .\ (\C{AccID}(r) = \C{ID}(t) \wedge \C{bal}(t) > \C{Amount}(t) \wedge \C{AccID}(r) = \C{ID}(t') \\ \wedge  \C{bal}(t') > \C{Amount}(t') \wedge t \xrightarrow{\F{ar}} t') \Rightarrow t \xrightarrow{\F{WW}} t'
\end{split}\\
\forall t,t'.\ t \xrightarrow{\F{RW}} t' \Rightarrow \neg (t' \xrightarrow{\F{vis}} t)\\
\forall t,t'.\ t \xrightarrow{\F{WW}} t' \Rightarrow t \xrightarrow{\F{vis}} t'
\end{gather}

We use $t,t'$ to denote invocations of the transaction, and $r$ to
denote a record in the database. We define the function \C{AccID} to
access the primary key of a record. Similarly, \C{ID}, \C{Amount},
etc. are functions which map an invocation to its parameters and local
variables. The existence of a dependence between two invocations
forces the existence of a record that both invocations must access, as
well as conditions on the local variables required to perform the
access (Eqn. 1). On the other hand, if two invocations are guaranteed
to write to the same location, there must exist a $\F{WW}$ dependency
between them (Eqn. 2). Now, it is not possible to have invocations
$t_1$ and $t_2$, obeying Eqns. (1)-(4) such that $t_1
\xrightarrow{\F{RW}} t_2$ and $t_2 \xrightarrow{\F{RW}} t_1$, the
condition necessary to induce a cycle and thus manifest a
serializability violation.

\begin{figure}[h]
\begin{tabular}{c c c}
$\xymatrix {\boldsymbol{t_1} \ar[r] \ar@/^2pc/[rrr] & t_2 \ar[r] & t_3 \ar[r] & \boldsymbol{t_4} \ar[r] & t_5} $ & & $\xymatrix {\boldsymbol{t_1} \ar[r] \ar@/^2pc/[rrrr] & t_2 \ar[r] & t_3 \ar[r] & t_4 \ar[r] & \boldsymbol{t_5}} $\\
\\
\\
$\xymatrix {t_1 \ar[r]  & \boldsymbol{t_2} \ar[r] \ar@/^2pc/[rr] & t_3 \ar[r] & \boldsymbol{t_4} \ar[r] & t_5} $ & & $\xymatrix {t_1 \ar[r] & \boldsymbol{t_2} \ar[r] \ar@/^2pc/[rrr] & t_3 \ar[r] & t_4 \ar[r] & \boldsymbol{t_5}} $
\end{tabular}
\caption{Different possibilities for paths of length 4 in the dependency graphs of the banking application. Note that transactions in bold perform writes.}
\label{Fig:3}
\vspace{-10pt}
\end{figure}
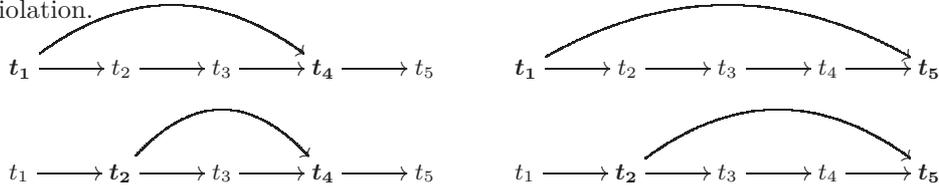
In fact, it is not possible to have a cycle of any arbitrary length in
a dependency graph of this application under PSI.  To show this, we
use the following observation : any long path in a dependency graph
generated by the above application will have chords in it, resulting
in a shorter path. In fact, it can be shown that the shortest path
between any two invocations in any dependency graph of the application
(if there is a path) will always be less than or equal to 3. This
can shown by using the above constraints (1)-(4) (and adding similar constraints for $\F{WR}$ edges) and then instantiating a path of
length 4 such that there is no chord between any of the nodes involved
in the path, and then showing the unsatisfiability of such an
encoding. Since a cycle is also a path, it is now sufficient to only
check for cycles of length 3, since any longer cycle will necessarily
induce a cycle of length less than or equal to 3.

Intuitively, this is happening in the banking application because the presence of any dependency edge between two nodes
implies that both invocations must access the same account, and at
least one of them must perform a write. Further, any two writes are
always related by a $\F{WW}$ edge. Now, as shown in Figure \ref{Fig:3}, in any path of length 4 in the dependency graph,
one of $t_1$ or $t_2$ and one of $t_4$ or $t_5$ must be a write,
which implies a chord between the two writes. Hence, there will always be a
shorter path of length less than or equal to 3 between $t_1$ and
$t_5$.

\section{Preliminaries}
\subsection{Input Language and Database Model}
\begin{mathpar}
\small
\begin{array}{lcl}
\multicolumn{3}{c}{
\C{v} \in \mathtt{Variables} \qquad \C{f} \in \mathtt{Fields} \qquad \mathcal{Q} \in \{\tt{MIN, MAX, COUNT}\}  } \\
\multicolumn{3}{c}{
\oplus \in \{+, -, \times, /\} \qquad \odot \in \{<, \leq, =, >, \geq\} \qquad \circ \in \{\wedge, \vee\}
}\\
e_d & \coloneqq & \C{f} \ALT \C{v} \ALT e_d \oplus e_d \ALT \mathbb{Z}\\
\phi_d & \coloneqq & \C{f} \odot e_d \ALT \C{f}\ \in \ \C{v} \ALT \neg \phi_d \ALT \phi_d \circ \phi_d\\ 
e_c & \coloneqq & \C{v} \ALT \C{CHOOSE } \C{v} \ALT e_c \oplus e_c  \ALT \mathbb{Z}\\
\phi_c & \coloneqq & \C{v} \odot e_c \ALT \C{v} = \C{NULL} \ALT \C{v}_1 \in \C{v}_2 \ALT \neg \phi_c \ALT \phi_c \circ \phi_c\\ 
c & \coloneqq & \tt{SELECT} \; \bar{f} \;\tt{AS}\; v\; \tt{WHERE}\; \phi_d \ALT \tt{SELECT}\; \mathcal{Q}\; f\; \tt{AS}\; v\; \tt{WHERE}\; \phi_d \ALT  \tt{UPDATE\; SET}\; f = e_c\; \tt{WHERE}\; \phi_d \ALT \\
& & \tt{INSERT\; VALUES}\; \bar{f} = \bar{e_c}\; \ALT \tt{DELETE\; WHERE}\; \phi_d \ALT v\; =\; e_c \ALT \tt{IF}\; \phi_c\; \tt{THEN}\; c\; \tt{ELSE}\; c \ALT c\; ; \;c \\ 
& & \tt{FOREACH}\; v_1\; \tt{IN}\; v_2\; \tt{DO}\; c\; \tt{END} \ALT \C{SKIP}\\
vlist & \coloneqq & \C{v} \ALT vlist,vlist\\
\mathcal{T} & \coloneqq & \mathtt{Tname}(vlist)\{\C{c}\}
\end{array}
\end{mathpar}
We start with description of the language of transactional programs in our framework. We assume a database model, where data is organized in tables with multiple records, where each record has multiple fields and transactions can insert/delete records and read/modify fields in selected records. The grammar is essentially a simplified version of standard SQL, allowing SQL statements which access the database to be combined with usual program connectives such as conditionals, sequencing and loops. Every transactional program $\mathcal{T}$ has a set of parameter variables ($vlist$) which are instantiated with values on invocation, and a set of local variables which are used to store intermediate values from the database (typically as output of \texttt{SELECT} queries). For a transactional program $\mathcal{T}$, let \C{Vars}$(\mathcal{T})$ be the set of parameters and local variables of $\mathcal{T}$. Let $\C{Stmts}(\mathcal{T})$ be the set of SQL statements (i.e. \C{INSERT}, \C{DELETE}, \C{SELECT} or \C{UPDATE}) in $\mathcal{T}$. 

To simplify the presentation, we will assume that there is only one table and each record is a set of values indexed by the set \C{Fields}. Furthermore all fields store integer values. The \C{FOREACH} loop iterates over a set of records in \C{v$_2$}, and assigns \C{v$_1$} to an individual record during each iteration. We call $\C{v}_2$ as the loop variable. Let $\mathcal{D}(\C{v})$ denote the nesting depth of $\C{v}$, which is 0 if \C{v} is assigned a value outside any loop (or is a parameter variable), and otherwise is the number of enclosing loops. For a variable \C{v} assigned a value inside a loop, let $\C{LVar}(\C{v},i)$ denote the loop variable at depth $i$, for all $1 \leq i \leq \mathcal{D}(\C{v})$. 

SQL statements use predicates $\phi_d$ to select records that would be accessed/modified, where $\phi_d$ allows all boolean combinations of comparison predicates between fields and values. Conditionals used inside \C{IF} statements ($\phi_c$) are only allowed to used local variables and parameters. To check whether the output of a \C{SELECT} query is empty, we use the conditional expression $\C{v} = \C{NULL}$, where \C{v} stores the output of the query. 

We assume a fixed non-empty subset of $\texttt{Fields}$ to be the primary key $\texttt{PK}$. Any two records must have distinct values in at least one of their $\texttt{PK}$ fields. Assume that there is a special field called $\C{Alive} \in \C{Fields}$ whose value is 1 if the record is in the database, 0 otherwise. Initially, all records are not \C{Alive}. When a record is inserted into the database, it becomes \C{Alive}, and when the record is deleted, it again becomes not \C{Alive}. 

\subsection{Abstract Executions}

Executions of transactional programs in our framework are expressed using an event structure, which is based on the approach used in \cite{BE16}.  The execution of a transaction instance consists of events, which are database operations. A database operation is a read or write to a field of a record. Let $\mathcal{R} = \texttt{PK} \rightarrow \mathbb{Z}$ be the set of all possible primary keys. Then, the set of all database operations is $ \mathcal{O} = \{\C{wri}(r,f,n)\ |\ r \in \mathcal{R}, f \in \texttt{Fields} \setminus \C{PK}, n \in \mathbb{Z}\} \cup \{\C{rd}(r,f,n)\ |\ r \in \mathcal{R}, f \in \texttt{Fields}, n \in \mathbb{Z}\}$ .

To simplify the presentation, we assume that a transaction reads (writes) at most once from (to) a field of a record and does not read any record that it writes, inserts or deletes. These assumptions allow us to ignore the ordering among events of a single transaction instance. Our approach can be easily adapted if these assumptions are not satisfied. 
\begin{definition}[Transaction Instance]
A transaction instance is a tuple $\sigma = (\C{TID}, \varepsilon)$, where $\C{TID}$ is a unique transaction instance-ID and $\varepsilon \subseteq \mathcal{O}$ is a set of events.
\end{definition}
In this work, we will assume that transactions are executed in an environment which guarantees atomicity and isolation (also called atomic visibility \cite{CE15}). That is, either all events of a transaction are made visible to other transactions, or none are, and the same set of transactions are visible to all events in a transaction. Atomicity and isolation are crucial properties for transactional programs, and both can be implemented efficiently in a replicated, distributed environment \cite{BU12,BA16}. Note that atomicity and isolation does not guarantee serializability, as seen in example in \S 2, and our goal is to explore serializability in this context of weak consistency. 
\begin{definition}[Abstract Execution]
An abstract execution is a tuple $\chi = (\Sigma, \F{vis}, \F{ar})$, where $\Sigma$ is a set of transaction instances, $\F{vis} \subseteq \Sigma \times \Sigma$ is an anti-symmetric, irreflexive relation, and $\F{ar} \subseteq \Sigma \times \Sigma$ is a total order on $\Sigma$ such that $\F{vis} \subseteq \F{ar}$.
\end{definition}

Intuitively, given transaction instances $\sigma, \sigma'$ in an abstract execution $\chi$, if $\sigma \xrightarrow{\F{vis}} \sigma'$, then all writes performed by $\sigma$ are visible to $\sigma'$ and hence may affect the output of the reads performed by $\sigma'$. $\F{ar}$ is used to order all writes to the same location. We use the notation $\sigma \vdash o$ to specify that transaction instance $\sigma$ performs a database operation $o$. The length of an abstract execution is defined to be the number of transaction instances involved in the execution (i.e. $|\Sigma|$).

Given a set of transaction instances $\Sigma'$, we use the notation $[\Sigma']_{<\C{wri}(r,f)>} = \{\sigma \in \Sigma'\ |\ \sigma \vdash \C{wri}(r,f,n), n \in \mathbb{Z} \}$ to denote the set of transactions which are writing to field \C{f} of record \C{r}. We use the notation $\F{MAX}_{\F{ar}}(\Sigma')$ to denote $\sigma \in \Sigma'$ such that $\forall \sigma' \in \Sigma'.\ \sigma = \sigma' \vee \sigma' \xrightarrow{\F{ar}} \sigma$. Given a transaction instance $\sigma$, we use $\F{vis}^{-1}(\sigma)$ to denote the set $\{\sigma' \in \Sigma\ |\ \sigma' \xrightarrow{\F{vis}} \sigma\}$. The last writer wins nature of the database dictates that a transaction reads the most recent value (according to $\F{ar}$) written by the transactions visible to it. Formally, this is specified as follows: $\sigma \vdash \C{rd}(r,f,n) \Rightarrow (f \not\in \C{PK} \Rightarrow \F{MAX}_{\F{ar}}([\F{vis}^{-1}(\sigma)]_{<\C{wri}(r,f)>}) \vdash \C{wri}(r,f,n)) \wedge (f \in \C{PK} \Rightarrow r(f) = n)$.

\begin{definition}[Dependency Graph]
Given an abstract execution $\chi=(\Sigma, \F{vis}, \F{ar})$, the dependency graph $G_{\chi} = (\Sigma, E)$ is a directed, edge-labeled multigraph where the edges and their labels are defined as follows : 
\begin{itemize}
\item $\sigma \xrightarrow{\F{WR}_{r,f}} \sigma'$ if $\sigma' \vdash \C{rd}(r,f,n)$ and $\sigma = \F{MAX}_\F{ar}([\F{vis}^{-1}(\sigma')]_{<\C{wri}(r,f)>})$.
\item $\sigma \xrightarrow{\F{WW}_{r,f}} \sigma'$ if $\sigma \vdash \C{wri}(r,f,n)$, $\sigma' \vdash \C{wri}(r,f,m)$ and $\sigma \xrightarrow{\F{ar}} \sigma'$.
\item $\sigma \xrightarrow {\F{RW}_{r,f}} \sigma'$ if $\sigma \vdash \C{rd}(r,f,n)$,  $\sigma' \vdash \C{wri}(r,f,m)$ and there exists another transaction instance $\sigma''$ such that $\sigma'' \xrightarrow{\F{WR}_{r,f}} \sigma$ and $\sigma'' \xrightarrow{\F{WW}_{r,f}} \sigma'$.
\end{itemize} 
\end{definition}
Edges in the dependency graph $G_{\chi}$ also induce corresponding binary relations on the transaction instances (we use the same notation for these relations). Let $\F{WR}, \F{WW}, \F{RW}$ be the union of $\F{WR}_{r,f}, \F{WW}_{r,f}, \F{RW}_{r,f}$ for all $r,f$ respectively. The following lemma follows directly from the definition:
\begin{lemma}
Given an abstract execution $\chi = (\Sigma, \F{vis}, \F{ar})$ and its dependency graph $G_{\chi} = (\Sigma, E)$, the following are true:
\begin{itemize}
\item If $\sigma \xrightarrow{\F{WR}_{r,f}} \sigma' \in E$, then $\sigma \xrightarrow{\F{vis}} \sigma'$.
\item If $\sigma \xrightarrow{\F{WW}_{r,f}} \sigma' \in E$, then $\sigma \xrightarrow{\F{ar}} \sigma'$.
\item If $\sigma \xrightarrow{\F{RW}_{r,f}} \sigma' \in E$, then $\neg (\sigma' \xrightarrow{\F{vis}} \sigma)$.
\end{itemize}
\end{lemma}
Note that all the proofs can be found in Appendix C. In our framework, transaction instances are generated by assigning values to all the parameter variables of a transactional program $\mathcal{T}$, written using the grammar specified in \S 3.1. We use the notation $\Gamma(\sigma)$ to denote the transactional program $\mathcal{T}$ associated with transaction instance $\sigma$. 

Different weak consistency and weak isolation models can be expressed by placing constraints on $\F{vis}$ and $\F{ar}$ relations associated with an abstract execution. This gives rise to the notion of \textbf{valid} abstract executions under a specific model, which are executions satisfying the constraints associated with those models. Below, we provide examples of several known weak consistency and weak isolation models:
\begin{itemize}
\item Full Serializability : $\Psi_{\mathrm{Ser}} \triangleq \F{vis} = \F{ar}$
\item Selective Serializability for transactional programs $\mathcal{T}_1, \mathcal{T}_2$ \cite{FE05b} : $\Psi_{\mathrm{Ser}(\mathcal{T}_1, \mathcal{T}_2)} \triangleq \forall \sigma_1, \sigma_2.$ $((\Gamma(\sigma_1) = \mathcal{T}_1 \wedge \Gamma(\sigma_2) = \mathcal{T}_2) \vee (\Gamma(\sigma_1) = \mathcal{T}_2 \wedge \Gamma(\sigma_2) = \mathcal{T}_1) \wedge \sigma_1 \xrightarrow{\F{ar}} \sigma_2) \Rightarrow \sigma_1 \xrightarrow{\F{vis}} \sigma_2$
\item Causal Consistency (CC) \cite{LL11} : $\Psi_{\mathrm{CC}} \triangleq \forall \sigma_1, \sigma_2, \sigma_3.$ $\sigma_1 \xrightarrow{\F{vis}} \sigma_2 \wedge \sigma_2 \xrightarrow{\F{vis}} \sigma_3 \Rightarrow \sigma_1 \xrightarrow{\F{vis}} \sigma_3$
\item Prefix Consistency (PC) (equivalent to Repeatable Read in centralized databases) \cite{TE13, BU15} : $\Psi_{\mathrm{PC}} \triangleq \forall \sigma_1, \sigma_2, \sigma_3.$ $\sigma_1 \xrightarrow{\F{ar}} \sigma_2 \wedge \sigma_2 \xrightarrow{\F{vis}} \sigma_3 \Rightarrow \sigma_1 \xrightarrow{\F{vis}} \sigma_3$
\item Parallel Snapshot Isolation \cite{SO11} : $\Psi_{\mathrm{PSI}} \triangleq \forall \sigma_1, \sigma_2.$ $\sigma_1 \xrightarrow{\F{WW}} \sigma_2 \Rightarrow \sigma_1 \xrightarrow{\F{vis}} \sigma_2$
\end{itemize}
Different models can be also be combined together to create a hybrid model. For example, $\Psi_{PSI} \wedge \Psi_{PC}$ is equivalent to Snapshot Isolation \cite{BE95} in centralized databases. Below, we formalize the classical notion of conflict serializability \cite{BE87} in our setting and then relate it to the presence of cycles in the dependency graph. 
\begin{definition}[Serializable Execution]
An abstract execution $\chi = (\Sigma, \F{vis}, \F{ar})$ is said to be serializable if there exists another abstract execution $\chi' = (\Sigma, \F{vis'}, \F{ar'})$ which satisfies $\Psi_{\mathrm{Ser}}$ such that $G_{\chi}$ and $G_{\chi'}$ are isomorphic.
\end{definition}
\begin{theorem}
Given an abstract execution $\chi = (\Sigma, \F{vis}, \F{ar})$, if there is no cycle in the dependency graph $G_{\chi}$, then $\chi$ is serializable.
\end{theorem} 

\subsection{Operational Semantics}
We now propose an operational semantics to generate abstract executions from transactional programs under a consistency specification. The purpose of the operational semantics is to link SQL statements with abstract database operations, and to prove the soundness of our encoding in FOL. Here, we only provide an informal overview and the full operational semantics can be found in Appendix B.

The semantics is a transition system $\mathcal{S}_{\mathbb{T}, \Psi} = (\Delta, \rightarrow)$ parametrized over a set of transactional programs $\mathbb{T}$ and a consistency specification $\Psi$. The state ($\delta \in \Delta$) is stored as a tuple $(\Sigma, \F{vis}, \F{ar}, \mathcal{P})$ where $\Sigma$ is the set of committed transaction instances, $\F{vis}$ and $\F{ar}$ are relations on $\Sigma$, and $\mathcal{P}$ is the running pool of transaction instances. The transitions are of two types : spawning a new instantiation of a transactional program $\mathcal{T} \in \mathbb{T}$ or executing a statement of a transaction instance in the running pool. When a new execution of a transaction instance begins, a subset of $\Sigma$ is non-deterministically selected to be made visible to the new instance. A view of the database is constructed for the new instance based on the set of visible transactions and the $\F{ar}$ relation (ensuring the last writer wins policy), and all queries of the transaction instance are answered on the basis of this view. At any point, any transaction instance from $\mathcal{P}$ can be non-deterministically selected for execution of its next statement. Any new event generated during the execution of a transaction instance is stored in the running pool. Finally, when a transaction instance wants to commit, it is checked whether the consistency specification ($\Psi$) is satisfied if the instance were to commit, and if yes, it is added to $\Sigma$. We can now define a valid abstract execution in terms of traces of the transition system:

\begin{definition}[Valid execution of $\mathbb{T}$ under $\Psi$]
An abstract execution $\chi = (\Sigma, \F{vis}, \F{ar})$ is said to be a valid execution produced by $\mathbb{T}$ under $\Psi$ if there exists a trace $(\{\}, \{\}, \{\},\{\}) \rightarrow^{*} (\Sigma, \F{vis}, \F{ar}, \{\})$ of the transition system $\mathcal{S}_{\mathbb{T},\Psi}$.
\end{definition}

\section{FOL Encoding}


\subsection{Vocabulary}

Given a set of transactional programs $\mathbb{T}$ and a consistency specification $\Psi$ we now show how to construct a formula in FOL such that any valid abstract execution $\chi$ of $\mathbb{T}$ under $\Psi$ and its dependency graph $G_{\chi}$ is a satisfying model of the formula. The encoding is parametric over $\mathbb{T}$ and $\Psi$. We first describe the vocabulary of the encoding. We define two uninterpreted sorts $\tau$ and $R$, such that members of $\tau$ are transaction instances, and members of $R$ are records. In addition, we also define a finite sort $\mathbb{T}$ which contains the transaction types, where each type is a transactional program. 

The function $\Gamma : \tau \rightarrow \mathbb{T}$ associates each transaction instance with its type. For each transactional program $\mathcal{T} \in \mathbb{T}$ and for each variable $\C{v} \in \C{Vars}(\mathcal{T})$, the variable projection function $\rho_{\C{v}}$ gives the value of \C{v} in a transaction instance. The signature of $\rho_{\C{v}}$ depends upon the type of the variable and whether it is assigned inside a loop. First, let us consider variables which are assigned values outside any loop. In our framework, variables are of two types : a value or a set of values. Further, the value can be either an integer (e.g. the parameter \C{ID} of the \C{withdraw} transaction) or a record. Let $\mathbb{V} = \mathbb{Z} \cup R$. If \C{v} is a value, the $\rho_{\C{v}}$ has the signature $\tau \rightarrow \mathbb{V}$. If \C{v} is a set of values, then $\rho_{\C{v}}$ is a predicate with signature $\mathbb{V} \times \tau \rightarrow \mathbb{B}$, such that $\rho_{\C{v}}(r,t)$ is true if $r$ belongs to \C{v} in the transaction instance $t$. 

Consider a loop of the form : \C{FOREACH}\; \C{v}$_1$\; \C{IN}\; \C{v}$_2$\; \C{DO}\; c\; \C{END}. All local variables which are assigned values inside the loop body (including \C{v}$_1$) will be indexed by values in the set \C{v}$_2$. Hence, if a local variable \C{v}$_3$ is assigned inside the loop, and it is a value, then $\rho_{\C{v}_3}$ will have the signature $\mathbb{V} \times \tau  \rightarrow \mathbb{V}$. On the other hand, if \C{v}$_3$ stores a set of values, then $\rho_{\C{v}_3}$ will have the signature $\mathbb{V} \times \mathbb{V} \times \tau  \rightarrow \mathbb{B}$, with the interpretation that $\rho_{\C{v}_3}(r_1, r_2, t)$ is true if $\C{v}_3$ contains $r_2$ in the iteration where $\C{v}_1$ is $r_1 \in \C{v}_2$. Similarly, nested loops will have local variables which are indexed by records in all enclosing loops. 

To summarize, the signature of $\rho_{\C{v}}$ is either $\mathbb{V}^{\mathcal{D}(\C{v})} \times \tau \rightarrow \mathbb{V}$ or $\mathbb{V}^{\mathcal{D}(\C{v}) + 1} \times \tau \rightarrow \mathbb{B}$. Similar to the variable projection function, the field projection function $\rho_{\C{f}} : R \rightarrow \mathbb{Z}$ is defined for each field $\C{f} \in \C{Fields}$, such that $\rho_{\C{f}}(r)$ gives the value of \C{f} in a record $r$. 

We define predicates $\F{WR}, \F{WW}, \F{RW}$ all of type $\tau \times \tau \rightarrow \mathbb{B}$ which specify the read, write and anti-dependency relations respectively between transaction instances.  We also define predicates $\F{WR}^{R}, \F{RW}^{R}, \F{WW}^{R}$ all of type $R \times \C{Fields} \times \tau \times \tau \rightarrow \mathbb{B}$ which provide more context by also specifying the records and fields causing the dependencies. Predicates $\F{vis}, \F{ar}$ of type $\tau \times \tau \rightarrow \mathbb{B}$ specify the visibility and arbitration relation between transaction instances. The predicate $\C{Alive} : R \times \tau \rightarrow \mathbb{B}$ indicates whether a record is \C{Alive} for a transaction instance. 

\subsection{Relating Dependences with Abstract executions}

By Lemma 4, in any abstract execution, the presence of a dependency edge between two transaction instances enforces constraints on the $\F{vis}$ and/or $\F{ar}$ relations between the two instances. The following formula encodes this along with basic constraints satisfied on $\F{vis}$ and $\F{ar}$:
\begin{equation}
\begin{split}
\varphi_{basic} & = \rulelabel{TotalOrder}(\F{ar}) \wedge \forall (t,s : \tau).\ (\F{vis}(t,s) \Rightarrow \neg \F{vis}(s, t))  \wedge (\F{vis}(t, s) \Rightarrow \F{ar}(t,s)) \\ & \wedge (\F{WR}(t,s) \Rightarrow \F{vis}(t,s)) \wedge (\F{WW}(t,s) \Rightarrow \F{ar}(t,s)) \wedge (\F{RW}(t,s) \Rightarrow \neg \F{vis}(s,t)) 
\end{split}
\end{equation}
The following formula encodes a fundamental constraint involving the dependency relations on the same field of the same record due to the last writer wins nature of the database:
\begin{equation*}
\begin{split}
\varphi_{dep} & = \bigwedge_{\C{f} \in \C{Fields}}\forall (t_1, t_2, t_3 : T) (r : R).\ \F{WR}^{R}(r, \C{f}, t_2, t_1) \wedge \F{RW}^{R}(r, \C{f}, t_1, t_3) \Rightarrow \F{WW}^{R}(r, \C{f}, t_2, t_3)
\end{split}
\end{equation*}
Finally, the consistency specification $\Psi$ can be directly encoded using the relations and functions defined in our vocabulary (we denote this formula by $\varphi_{\Psi}$). 

\subsection{Relating dependences with transactional programs}

The presence of a dependency edge between two transaction instances places constraints on the type of transactional programs generating the instances and their parameters. To automatically infer these constraints, we use the following strategy : if there is a dependency edge between two instances, then there must exist SQL statements in both transactions which access a common record. 

To encode this, we first extract the conditions under which a SQL statement in a transactional program can be executed. By performing a simple syntactic analysis over the code of a transaction $\M{T}$, we obtain a mapping $\Lambda_{\M{T}}$ from each SQL statement in \C{Stmts}($\mathcal{T}$) to a conjunction of enclosing $\C{IF}$ conditionals (the complete algorithm can be found in Appendix A).
\begin{figure}[h]
\begin{mathpar}
\small
\begin{array}{lclr}
\llbracket \C{v} = \C{NULL} \rrbracket_{t} & = & (\exists (r_1,\ldots, r_{\mathcal{D}(\C{v})} : R).\ \bigwedge_{i=1}^{\mathcal{D}(\C{v})} \mathcal{V} (\llbracket r_i \in \C{LVar}(\C{v},i) \rrbracket_{t}), &  \F{fresh}(r_1, \ldots, r_{\mathcal{D}(\C{v})}, r)\\
 & & \forall (r : R).\neg \rho_{\C{v}}(r_1, \ldots, r_{\mathcal{D}(\C{v})}, r, t))\\
\llbracket r \in \C{v} \rrbracket_{t} & = & (\exists (r_1,r_2,\ldots, r_{\mathcal{D}(\C{v})} : R).\  \bigwedge_{i=1}^{\mathcal{D}(\C{v})} \mathcal{V} (\llbracket r_i \in \C{LVar}(\C{v},i) \rrbracket_{t}), & \F{fresh}(r_1, \ldots, r_{\mathcal{D}(\C{v})})\\
 & & \rho_{\C{v}}(r_1, \ldots, r_{\mathcal{D}(\C{v})}, r, t)) \\
\llbracket \C{v}_1 \in \C{v}_2 \rrbracket_{t} & = & (\varphi_1 \wedge \varphi_2, \psi_2) &  \llbracket \C{v}_1 \rrbracket_{t} = (\varphi_1, \psi_1)\\
& & & \llbracket \psi_1 \in \C{v}_2 \rrbracket_{t} = (\varphi_2, \psi_2)\\
\llbracket \C{f} \odot e \rrbracket_{t,r} & = & \begin{cases} (\varphi, \rho_{\C{f}}(r) \odot \psi) & \text{if } \C{f} \cup \mathcal{F}(e) \subseteq \C{PK}  \\ (\F{true}, \F{true}) & \text{otherwise} \end{cases} \\
& & & \llbracket e \rrbracket_{t,r} = (\varphi, \psi)\\
\end{array}
\end{mathpar}
\caption{Encoding conditionals and \C{WHERE} clauses}
\label{Fig:conds}
\vspace{-10pt}
\end{figure}

The FOL encoding of all conditionals in a program and all \C{WHERE} clauses in a SQL statement is constructed by replacing variables and fields with the corresponding variable projection and field projection functions respectively. A representative set of rules for this encoding are shown in Fig. \ref{Fig:conds}. For conditionals $\phi$ used in \C{IF} statements, we use the notation $\llbracket \phi \rrbracket_{t}$ to describe the FOL encoding specialized to transaction instance $t$. The interpretation is that $\llbracket \phi \rrbracket_{t}$ is satisfiable only if the conditional $\phi$ is true in the transaction instance $t$. If $\phi$ is inside a loop, then $\llbracket \phi \rrbracket_{t}$ must be satisfiable if $\phi$ is true in any arbitrary iteration of the enclosing loop(s) in $t$. For this reason, $\llbracket \phi \rrbracket_{t}$ is actually represented as a tuple $(\varphi, \psi)$, where $\varphi$ chooses any arbitrary iteration of enclosing loops, and the formula $\psi$ is the value of the conditional in that iteration. We define an evaluation function $\mathcal{V}(\varphi, \psi) = \varphi \wedge \psi$ which gives the final FOL encoding.

The formula $\varphi$ chooses an iteration by instantiating records belonging to loop variables of all enclosing loops. For example, consider the encoding of $\C{v} = \C{NULL}$. Here, $\varphi$ instantiates a record belonging to the loop variable of every enclosing loop of \C{v} (encoded as $\mathcal{V} (\llbracket r_i \in \C{LVar}(\C{v},i) \rrbracket_{t})$), and $\psi$ encodes that $\rho_{\C{v}}$ in the chosen iteration is $\F{false}$ for every record. Similarly, in the encoding of $\llbracket r \in \C{v} \rrbracket_{t}$, $\rho_{\C{v}}$ must be $\F{true}$ for the record $r$. In the encoding of $\llbracket \C{v}_1 \in \C{v}_2 \rrbracket_{t}$, we first obtain the value of $\C{v}_1$ (the second term in the tuple $\llbracket \C{v}_1 \rrbracket_{t}$), and then check whether it is present in $\C{v}_2$.  


A similar procedure is used to obtain the encoding of the \C{WHERE} clauses used inside SQL statements. Since \C{WHERE} clauses are evaluated on records, the encoding is specialized on both records and transaction instances, for which we use the notation $\llbracket \phi \rrbracket_{t,r}$. The interpretation is that $\llbracket \phi \rrbracket_{t,r}$ is satisfiable only if $\phi$ is $\F{true}$ for transaction instance $t$ on record $r$. The encoding replaces the field accesses with the corresponding field projection function applied on $r$. Note that the field projection function is only used for primary key fields which are accessed within \C{WHERE} clauses (expressed as $\M{F} \subseteq \C{PK}$). The complete encoding for all types of conditionals and \C{WHERE} clauses can be found in the Appendix A.

As stated earlier, our strategy is to encode the necessary condition for a dependency edge based on the access of a common record. For each pair of transaction types $\mathcal{T}_1, \mathcal{T}_2 \in \mathbb{T}$, each dependency type $\mathcal{R} \in \{\F{WR}, \F{RW}, \F{WW}\}$, and each pair of SQL statements $\C{c}_1 \in \C{Stmts}(\mathcal{T}_1), \C{c}_2 \in \C{Stmts}(\mathcal{T}_2)$, we compute a necessary condition $\eta_{\C{c}_1, \C{c}_2}^{\mathcal{R}\rightarrow,\mathcal{T}_1, \mathcal{T}_2}(t_1, t_2)$ for dependency $\mathcal{R}$ to exist between instances $t_1$ and $t_2$ of types $\mathcal{T}_1$ and $\mathcal{T}_2$ due to statements $\C{c}_1$ and $\C{c}_2$ respectively. The following formula encodes the fact that a dependency between two transaction instances can be caused due to a dependency between any two SQL statements in those transactions:
\begin{equation*}
\varphi_{\mathcal{R}\rightarrow, \mathcal{T}_1, \mathcal{T}_2} \triangleq \forall (t_1, t_2 : \tau). (\Gamma(t_1) = \mathcal{T}_1 \wedge \Gamma(t_2) = \mathcal{T}_2 \wedge \mathcal{R}(t_1, t_2)) \Rightarrow \bigvee_{\substack{\C{c}_1 \in \C{Stmts}(\mathcal{T}_1)\\ \C{c}_2 \in \C{Stmts}(\mathcal{T}_2)}}\eta_{\C{c}_1, \C{c}_2}^{\mathcal{R}\rightarrow,\mathcal{T}_1, \mathcal{T}_2}(t_1, t_2)
\end{equation*}
The general format of $\eta_{\C{c}_1, \C{c}_2}^{\mathcal{R}\rightarrow,\mathcal{T}_1, \mathcal{T}_2}(t_1, t_2)$ is following : it is the conjunction of the conditionals required to execute the statements $\C{c}_1$ and $\C{c}_2$ (i.e. $\Lambda_{\mathcal{T}_1}(\C{c}_1)$ and $\Lambda_{\mathcal{T}_2}(\C{c}_2)$) in $t_1$ and $t_2$ resp. and the \C{WHERE} clauses of the two statements evaluated on some record $r$. If they can never access the same field of the same record, then $\eta_{\C{c}_1, \C{c}_2}^{\mathcal{R}\rightarrow,\mathcal{T}_1, \mathcal{T}_2}(t_1, t_2)$ is simply $\F{false}$. While this is the general format of the clauses, in addition, we can also infer more information depending upon the type of the SQL statements. To illustrate this we present a sample rule below:
$$
\RULE{\C{c}_1 \equiv  \tt{SELECT\ MAX} \; \F{f} \;\tt{AS}\; \C{v}\; \tt{WHERE}\; \phi_1 \quad \C{c}_2 \equiv \tt{UPDATE\; SET}\; \F{f} = e\; \tt{WHERE}\; \phi_2 \\ \C{c}_1 \in  \C{Stmts}(\mathcal{T}_1) \quad \C{c}_2 \in  \C{Stmts}(\mathcal{T}_2) \quad \Gamma(t_1) = \mathcal{T}_1 \quad \Gamma(t_2) = \mathcal{T}_2 \quad  \llbracket \C{v} \rrbracket_{t_1} = (\varphi_1, \psi_1) \quad \llbracket \C{e} \rrbracket_{t_2} = (\varphi_2, \psi_2)}
{\eta_{\C{c}_1, \C{c}_2}^{\F{RW}\rightarrow,\mathcal{T}_1, \mathcal{T}_2}(t_1, t_2) = (\exists r.\ \mathcal{V}(\llbracket \Lambda_{\mathcal{T}_1}(\C{c}_1) \rrbracket_{t_1}) \wedge \mathcal{V}(\llbracket \phi_1 \rrbracket_{t_1, r} ) \wedge  \mathcal{V}(\llbracket \Lambda_{\mathcal{T}_2}(\C{c}_2) \rrbracket_{t_2}) \wedge \mathcal{V}(\llbracket \phi_2 \rrbracket_{t_2, r}) \wedge \\ \C{Alive}(r, t_2) \wedge  \varphi_1 \wedge \varphi_2 \wedge \psi_1 < \psi_2 )}
$$

The rule encodes a necessary condition for an anti-dependency to exist from a \C{SELECT MAX} to a \C{UPDATE} statement. First, it encodes that the conflicting SQL statements actually execute in their respective transactions and there is a common record which satisfies the \C{WHERE} clauses of both statements. \C{SELECT MAX} selects the record with the maximum value in the field \C{f} among all records that satisfy $\phi_1$, and stores the value in variable \C{v}. If there is an anti-dependency from \C{SELECT MAX} to \C{UPDATE}, then the updated value must be greater than output of \C{SELECT MAX}, because otherwise, the update does not affect the output of \C{SELECT MAX}. The complete set of rules can be found in the Appendix A.

In addition, some transaction instances may be guaranteed to execute certain SQL statements, which forces the presence of a dependency edge between them. For example, if two transaction instances are guaranteed to update the same field of a record, then there must be a $\F{WW}$ dependeny between them. For each pair of transaction types $\mathcal{T}_1, \mathcal{T}_2 \in \mathbb{T}$, each dependency type $\mathcal{R} \in \{\F{WR}, \F{RW}, \F{WW}\}$, and each pair of SQL statements $\C{c}_1 \in \C{Stmts}(\mathcal{T}_1), \C{c}_2 \in \C{Stmts}(\mathcal{T}_2)$, we compute a condition $\eta_{\C{c}_1, \C{c}_2}^{\rightarrow\mathcal{R},\mathcal{T}_1, \mathcal{T}_2}(t_1, t_2)$  which forces the dependency $\mathcal{R}$ to exist between instances $t_1$ and $t_2$ of types $\mathcal{T}_1$ and $\mathcal{T}_2$ respectively due to $\C{c}_1$ and $\C{c}_2$. The following formula encodes this:
\begin{gather*}
\varphi_{\rightarrow\mathcal{R}, \mathcal{T}_1, \mathcal{T}_2} \triangleq \forall t_1, t_2. (\Gamma(t_1) = \mathcal{T}_1 \wedge \Gamma(t_2) = \mathcal{T}_2 \wedge \bigvee_{\substack{\C{c}_1 \in \C{Stmts}(\mathcal{T}_1)\\ \C{c}_2 \in \C{Stmts}(\mathcal{T}_2)}}\eta_{\C{c}_1, \C{c}_2}^{\rightarrow\mathcal{R},\mathcal{T}_1, \mathcal{T}_2}(t_1, t_2))  \Rightarrow \mathcal{R}(t_1, t_2)\\
\RULE{\C{c}_1 \equiv \tt{UPDATE\; SET}\; f = e_1\; \tt{WHERE}\; \phi_1 \quad \C{c}_2 \equiv \tt{UPDATE\; SET}\; f = e_2\; \tt{WHERE}\; \phi_2 \\ \C{c}_1 \in  \C{Stmts}(\mathcal{T}_1) \quad \C{c}_2 \in  \C{Stmts}(\mathcal{T}_2) \quad \Gamma(t_1) = \mathcal{T}_1 \quad \Gamma(t_2) = \mathcal{T}_2}
{\eta_{\C{c}_1, \C{c}_2}^{\rightarrow\F{WW},\mathcal{T}_1, \mathcal{T}_2}(t_1, t_2) =  (\exists r.\ \mathcal{V}(\llbracket \Lambda_{\mathcal{T}_1}(\C{c}_1) \rrbracket_{t_1}) \wedge \mathcal{V}(\llbracket \phi_1 \rrbracket_{t_1, r} ) \wedge  \mathcal{V}(\llbracket  \Lambda_{\mathcal{T}_1}(\C{c}_1) \rrbracket_{t_2}) \wedge \mathcal{V}(\llbracket \phi_2 \rrbracket_{t_2, r}) \wedge \\ \C{Alive}(r, t_1) \wedge \C{Alive}(r, t_2) \wedge \F{ar}(t_1, t_2)) }
\end{gather*}

$\eta_{\C{c}_1, \C{c}_2}^{\rightarrow\mathcal{R},\mathcal{T}_1, \mathcal{T}_2}(t_1, t_2)$ is computed in the same manner as $\eta_{\C{c}_1, \C{c}_2}^{\mathcal{R},\mathcal{T}_1, \mathcal{T}_2\rightarrow}(t_1, t_2)$. As an example consider the above rule. Two \C{UPDATE} statements modifying the same field are guaranteed to cause a $\F{WW}$ dependency if both statements actually execute in their respective transactions, and there exists a common record accessed by both statements which is \C{Alive} to both transactions. 

In addition, there are some auxiliary facts which are satisfied by all abstract executions (which we encode as the formula $\varphi_{aux}$) such as a record present in the output variable of a \C{SELECT} query must satisfy the \C{WHERE} clause of the query, the value of the iterator variable in a loop must belong to the loop variable, etc. For more details, we again refer to the Appendix. The final encoding is defined as follows:
\begin{equation}
\varphi_{\mathbb{T}, \Psi} \triangleq \varphi_{basic} \wedge \varphi_{dep} \wedge \bigwedge_{\mathcal{R} \in \{\F{WR}, \F{RW}, \F{WW}\}}\bigwedge_{\mathcal{T}_1, \mathcal{T}_2 \in \mathbb{T}} (\varphi_{\mathcal{R} \rightarrow, \mathcal{T}_1, \mathcal{T}_2} \wedge \varphi_{\rightarrow \mathcal{R}, \mathcal{T}_1, \mathcal{T}_2}) \wedge \varphi_{\Psi} \wedge \varphi_{aux}
\end{equation}
\begin{theorem}
Given a set of transactional programs $\mathbb{T}$ and a consistency specification $\Psi$, for any valid abstract execution $\chi = (\Sigma, \F{vis}, \F{ar})$ generated by $\mathbb{T}$ under $\Psi$ and its dependency graph $G_{\chi}$, there exists a satisfying model of the formula $\varphi_{\mathbb{T}, \Psi}$ with $\tau = \Sigma$ and the binary predicates $\F{vis}, \F{ar}, \F{WR}, \F{RW}, \F{WW}$ being equal to the corresponding relations in $\chi$ and $G_{\chi}$.
\end{theorem} 
Note that $\varphi_{\mathbb{T}, \Psi}$ is always satisfiable, since the empty abstract execution is a satisfying model. In the next section, we will ask for non-empty but bounded satisfying models of $\varphi_{\mathbb{T}, \Psi}$ with specific properties.
\section{Applications}

\subsection{Bounded Anomaly Detection}
By Theorem 6, any execution which violates serializability must have a cycle in its dependency graph. We can directly instantiate a dependency graph which contains a cycle of bounded length and then ask for a satisfying model of the formula built in the previous section which contains the cycle. We introduce a new predicate $D : \tau \times \tau \rightarrow \mathbb{B}$ which represents the presence of any dependency edge between two transaction instances : $\varphi_{D} \triangleq \forall (t_1, t_2 : \tau). \F{D}(t_1, t_2) \Leftrightarrow (t_1 = t_2) \vee \F{WR}(t_1, t_2) \vee \F{RW}(t_1, t_2) \vee \F{WW}(t_1, t_2)$. A cycle of length less than or equal to $k$ can now be directly encoded as follows: $\varphi_{\mathrm{Cycle,k}} \triangleq \exists t_1, \ldots, t_k.\ \bigwedge_{i=1}^{k-1} \F{D}(t_i, t_{i+1}) \wedge \F{D}(t_k, t_1)  \wedge (t_1 \neq t_k)$.
\begin{theorem}
Given a set of transactional programs $\mathbb{T}$ and a consistency specification $\Psi$, if $\varphi_{\mathbb{T}, \Psi} \wedge \varphi_{D} \wedge \varphi_{\mathrm{Cycle,k}}$ is $\F{UNSAT}$, then all valid abstract executions produced by $\mathbb{T}$ under $\Psi$ of length less than or equal to $\F{k}$ are serializable.
\end{theorem}

\subsection{Verifying Serializability : The Shortest Path Approach}
We propose a condition, which can be also be encoded in FOL, and which if satisfied would imply that it is enough to check for violations of bounded length to prove the absence of violations of any arbitrary length. 

The condition is based on the simple observation that any long path in the dependency graph could induce a short path due to chords among the nodes in the path (as demonstrated in the example in \S 2). This would imply that any long cycle would also induce a short cycle, and hence lack of short cycles would imply the lack of longer cycles. To check for this condition, we encode a shortest path of length $k$ in the dependency graph and then ask whether there is a satisfying model:
\begin{equation*}
\varphi_{\mathrm{Shortest\ Path,k}} \triangleq \exists t_1, \ldots, t_k, t_{k+1}.\ \bigwedge_{i=1}^{k} \F{D}(t_i, t_{i+1}) \wedge \bigwedge_{i=1}^{k-1} \bigwedge_{j=i+2}^{k+1} \neg \F{D}(t_i, t_j) \wedge \bigwedge_{\substack {1 \leq i < j \leq k+1}} t_i \neq t_j
\end{equation*}
The condition instantiates a path of length $k$ in the dependency graph and also asserts the absence of any chord, which implies that the path is shortest. If there does not exist a shortest path of length $k$, then there also cannot exist a shortest path of greater length, because if not, such a path would necessarily contain a shortest path of length $k$. Now, it is enough to check for cycles of length less than or equal to $k$, because any longer cycle would contain a path of length at least $k$, which would imply the presence of a shorter path and thus a cycle of length less than or equal to $k$.
\begin{theorem}
Given a set of transactional programs $\mathbb{T}$ and a consistency specification $\Psi$, if both $\varphi_{\mathbb{T}, \Psi} \wedge \varphi_{D} \wedge \varphi_{\mathrm{Shortest\ Path,k}}$ and $\varphi_{\mathbb{T}, \Psi} \wedge \varphi_{D} \wedge \varphi_{\mathrm{Cycle,k}}$ are $\F{UNSAT}$, then all valid abstract executions produced by $\mathbb{T}$ under $\Psi$ are serializable. 
\end{theorem}
\subsection{Verifying Serializability : An Inductive Approach}
We now present an alternative approach to verifying serializability which uses the transitivity and irreflexivity of the $\F{ar}$ relation to show lack of cycles. In this approach, our goal is to show that if there is a path in the dependency graph from $t_1$ to $t_2$, then $t_1 \xrightarrow{\F{ar}} t_2$. By the irreflexivity of $\F{ar}$, this would imply that there cannot be a cycle in the dependency graph. Since paths can be of arbitrary length, we will use the transitivity of $\F{ar}$ and an inductive argument to obtain a simple condition which can be encoded in FOL.
\begin{lemma}
Given a set of transactional programs $\mathbb{T}$, a consistency specification $\Psi$ and a subset of programs $\mathbb{T}' \subseteq \mathbb{T}$, \textbf{if} for all valid executions $\chi$ and their dependency graphs $G_{\chi}$, the following conditions hold:
\begin{enumerate}
\item if $\sigma_1 \rightarrow \sigma_2$ in $G_{\chi}$ and $\Gamma(\sigma_1) \in \mathbb{T'}$, then $\sigma_1 \xrightarrow{\F{ar}} \sigma_2$
\item if $\sigma_1 \rightarrow \sigma_2 \rightarrow \sigma_3$ in $G_{\chi}$, then either $\sigma_1 \xrightarrow{\F{ar}} \sigma_3$ or $\sigma_2 \xrightarrow{\F{ar}} \sigma_3$
\end{enumerate}
\textbf{then} all valid executions which contain at least one instance of a program in $\mathbb{T}'$ are serializable.
\end{lemma}  
The proof uses an inductive argument to show that if there is path from $\sigma_1$, an instance of a program in $\mathbb{T}'$ to any other instance $\sigma_2$, then $\sigma_1 \xrightarrow{\F{ar}} \sigma_2$. This would imply that any instance of $\mathbb{T}'$ cannot be present in a cycle. The above conditions can be directly encoded in FOL:
\begin{equation}
\begin{split}
\varphi_{\mathrm{Inductive}, \mathbb{T}'} \triangleq (\exists (t_1, t_2 : \tau).\ \Gamma(t_1) \in \mathbb{T}' \wedge \F{D}(t_1, t_2) \wedge t_1 \neq t_2 \wedge \neg \F{ar}(t_1, t_2)) \vee \\
(\exists(t_1, t_2, t_3 : \tau). \F{D}(t_1, t_2) \wedge \F{D}(t_2, t_3) \wedge \bigwedge_{\substack{1 \leq i < j \leq 3}} t_i \neq t_j \wedge \neg \F{ar}(t_1, t_3) \wedge \neg \F{ar}(t_2, t_3) )
\end{split}
\end{equation}
\begin{theorem}
Given a set of programs $\mathbb{T}$ and a consistency specification $\Psi$, if $\varphi_{\mathbb{T}, \Psi} \wedge \varphi_{D} \wedge \varphi_{\mathrm{Inductive}, \mathbb{T}'}$ is $\F{UNSAT}$, then all valid executions of $\mathbb{T}$ under $\Psi$ which contains at least one instance of a program in $\mathbb{T}'$ are serializable.
 \end{theorem}
If $\mathbb{T'} = \mathbb{T}$, then all valid executions of $\mathbb{T}$ are serializable, otherwise, we can focus only on programs in $\mathbb{T} \setminus \mathbb{T}'$, and re-apply the technique with $\varphi_{\mathbb{T'}, \Psi} \wedge \varphi_{D} \wedge \varphi_{\mathrm{Inductive}, \mathbb{T}''}$ for $\mathbb{T''} \subseteq \mathbb{T'}$. In the next section, we show how we use this technique to verify serializability of TPC-C, a real-world database benchmark.
\section{Case Studies}

We have developed a tool called \rulelabel{Anode} which takes a set of programs written in the language presented in \S 3.1 and a consistency specification and uses the encoding rules presented in \S 4 to automatically generate an FOL encoding. We use the Z3 SMT solver to determine the satisfiabiliy of the generated formulaes. In order to evaluate the effectiveness of our approach, we have applied the proposed technique on TPC-C \cite{TPCC}, a well-known Online Transaction Processing (OLTP) benchmark widely used in the database community, and a Courseware application (used in \cite{GO16}) which is a representative of course registration systems used in universities.

\textbf{TPC-C} : TPC-C has a complex database schema with 9 tables, and complex application logic in its 5 transactions. The transactions contain loops and conditionals, have multiple parameters and behave differently depending upon the values of the parameters; they also use complex queries such as \C{SELECT MIN} and \C{SELECT MAX}. To the best of our knowledge, this is the first automated static analysis for validating serializability of TPC-C under weak consistency. 

\begin{wrapfigure}{r}{0.5 \textwidth}
\small
\vspace{-10pt}
$
\xymatrix@R=1.5pt{& *+[F]{\C{Order-Status}_1} \ar[rd]^{\F{RW}}\\
 *+[F]{\C{New-Order}} \ar[ru]^{\F{WR}}   & &   *+[F]{\C{Payment}} \ar[ld]^{\F{WR}} \\
 & *+[F]{\C{Order-Status}_2} \ar[lu]^{\F{RW}}
 }
$
\caption{Long fork anomaly in TPC-C under PSI}
\label{Fig:anomaly}
\vspace{-20pt}
\end{wrapfigure}
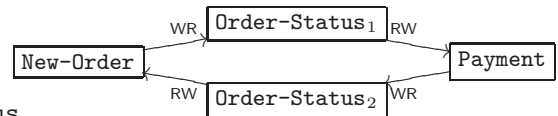

Under eventual consistency, TPC-C has a number of `lost update' anomalies, similar to the anomaly in the banking application described in \S 2. These anomalies are small in length and were automatically detected using encoding presented in \S 5.1 (with $k=2$) . To get rid of these anomalies, we upgraded the consistency specification to PSI \cite{SO11}. Under PSI, we did not find any anomalies for $k=2$ or $k=3$, but for $k=4$, the `long fork' anomaly involving the New-Order, Payment and Order-Status transactions was discovered, as shown in Fig. \ref{Fig:anomaly}.

This anomaly happens because the \C{New-Order} and \C{Payment} transactions update two different tables (\C{Order} and \C{Customer} table resp.) while the \C{Order-Status} transaction reads both those tables. Since there is no synchronization between \C{New-Order} and \C{Payment} transactions, it is possible for $\C{Order-Status}_1$ to see the update of \C{New-Order} but not \C{Payment}, and the vice versa for $\C{Order-Status}_2$. We also discovered a similar anomaly involving two instances of \C{New-Order} and two instances of \C{Stock-level} transactions.

To get rid of these anomalies, we further upgraded the consistency level to Snapshot Isolation (SI), after which we did not find any anomalies for $k=4$. We then turned our attention to verifying serializability of TPC-C under SI. We first tried the Shortest Path approach (which worked well for the banking application), but we were able to discover a long path (which can be arbitrarily extended) without any chords. Next, we tried the inductive approach, which was successful in proving serializability of TPC-C. Specifically, with $\mathbb{T}' = \{\C{New-Order}, \C{Payment}\}$, the formula $\varphi_{\mathrm{Inductive}, \mathbb{T}'}$ was shown to be $\F{UNSAT}$, and with the remaining 3 transactions $\varphi_{\mathrm{Inductive}, \{\C{Delivery}\}}$ was $\F{UNSAT}$. The remaining two transactions do not have any dependencies between them, which implies that all executions of TPC-C under SI are serializable. 

\textbf{Courseware} : The Courseware application maintains a database of courses and students, and provides the functionality of adding/removing students and courses, and enrolling students into courses subject to course capacities. Under EC, the following anomalies were discovered by our encoding : (1) two concurrent \C{Enroll} transactions may enroll students beyond the course capacity, (2) two courses with the same name or two students with the same name may be registered, (3) a student may be enrolled in a course which is being concurrently removed, or the student is being concurrently removed. Note that all these anomalies were discovered for $k=2$.

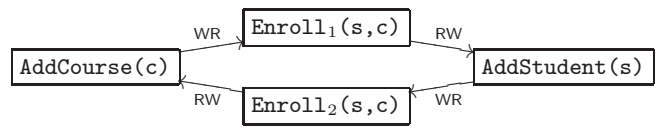
\begin{wrapfigure}{r}{0.6 \textwidth}
\small
\vspace{-10pt}
$
\xymatrix@R=1.5pt{& *+[F]{\C{Enroll}_1\C{(s,c)}} \ar[rd]^{\F{RW}}\\
 *+[F]{\C{AddCourse(c)}} \ar[ru]^{\F{WR}}   & &   *+[F]{\C{AddStudent(s)}} \ar[ld]^{\F{WR}} \\
 & *+[F]{\C{Enroll}_2\C{(s,c)}} \ar[lu]^{\F{RW}}
 }
$
\caption{Long fork anomaly in the Courseware application under PSI}
\label{Fig:anomaly-courseware}
\vspace{-20pt}
\end{wrapfigure}

In order to remove these violations, we upgraded the consistency model in a number of ways : the \C{Enroll} transaction was upgraded to PSI, while selective serializability was used for two instances of \C{AddCourse} and \C{AddStudent}, and for instances of \C{Enroll} and \C{RemCourse}, \C{Enroll} and \C{RemStudent}. While these upgrades took care of the above mentioned anomalies, we discovered a new long fork anomaly (for $k=4$) as shown in Fig. \ref{Fig:anomaly-courseware}. Here, two \C{Enrol}l transactions trying to enroll a student (\C{s}) into a course (\C{c}) see conflicting views of the database, with one \C{Enroll} witnessing the student but not the course, and vice versa for the other. We note that while this is an actual serializability violation, it is completely harmless as both enroll transactions which witness inconsistent database states will fail, so that final database state is the same as that which manifests at the end of an execution in which neither of the two enrolls happen. This is a limitation of our analysis as it does not provide any way to ignore harmless serializability violations. We plan to address this issue in future works. 

In order to remove this violation, we upgraded the consistency level of \C{Enroll} to SI, after which we did not find any anomalies. Next, we moved to verification, and here we were successfully able to use the Shortest Path approach and prove that there does not exist a shortest path in any dependency graph of the Courseware application of length greater than or equal to 8. Along with the fact there does not exist any cycle of length less than or equal to 8, this implies that any execution of the application is serializable. 	Note that in all instances, the solver produced its output in few (< 10) seconds.
\section{Related Work and Conclusion}

Serializability is a well-studied problem in the database community, but there is a lack of static automated techniques to check for serializability of database applications. Early work by Fekete et. al. \cite{FE05a} and Jorwekar et. al. \cite{JO07} proposed lightweight syntactic analyses to check for serializability under SI in centralized databases, by looking for dangerous structures in the static dependency graph of an application (which is an over-approximation of all possible dynamic dependency graphs). Several recent works \cite{BE16, CE15, CE16, CE17, ZH13, WA17} have continued along this line, by deriving different types of dangerous structures in dependency graphs that are possible under different weak consistency mechanisms, and then checking for these structures on static dependency graphs. 

However, static dependency graphs are highly imprecise representations of actual executions, and any analysis reliant on these graphs is likely to yield a large number of false positives. Indeed, several works \cite{BE16, CE16, CE17} recognize this and propose complex conditions to reduce false positives for specific consistency mechanisms, but these works do not provide any automated methodology to check those conditions on actual programs. Further, application logic could prevent these harmful structures from manifesting in actual executions, for example as in TPC-C, which has a harmful structure in its static dependency graph under SI, but which does not appear in any dynamic dependency graph. In our work, we precisely model the application logic and the consistency specification using FOL, so that the solver would automatically derive harmful structures which are possible under the given consistency specification and search for them in actual dependency graphs taking application logic into account. 

\cite{BR18} proposes a static analysis for serializability under causal consistency by constructing actual dependency graphs with cycles using a FOL encoding. While this work is similar to ours in spirit, there are several key points of differences : their notion of serializability is stronger than ours, since they allow transactions to be grouped together in sessions, with the serial order forced to accommodate the session order. While this eases the task of verifying serializability for unbounded executions, it also results in a large number of harmless serializability violations (for which they propose various ad hoc filtering approaches). Further, their focus is on programs operating on high-level data types rather than SQL programs, and their analysis is not parametric on consistency specifications.

There are also dynamic anomaly detection techniques \cite{ZE14, CA09, BR17} which either build the dependency graphs at run-time and check for cycles, or analyze the trace of events after execution. These approaches do not provide any guarantee that all anomalies will be detected, even for bounded executions. A number of approaches have been proposed recently \cite{SI15, GO16, KA18, CR17} which attempt to verify that high-level application invariants are preserved under weak consistency. These approaches are also parametric on consistency specifications, but they are not completely automated as they require correctness conditions in the form of invariants from the user, and they do not tackle serializability.  

To conclude, in this paper we take the first step towards building a precise, fully automated static analysis for serializability of database applications under weak consistency. We leverage the acyclic dependency graph based characterization of serializability and the framework of abstract executions to develop a FOL based analysis which is parametric on the consistency specification. We show how our approach can be used to detect bounded anomalies, and to verify serializability under specific conditions for unbounded executions. We show the practicality of our approach by successfully applying it on a real-world database benchmark.

\bibliography{db}

\begin{thebibliography}{10}

\bibitem{TPCC}
Tpc-c benchmark.
\newblock
  \url{http://www.tpc.org/tpc_documents_current_versions/pdf/tpc-c_v5.11.0.pdf}.
\newblock Online; Accessed 20 April 2018.

\bibitem{AD00}
Atul Adya, Barbara Liskov, and Patrick~E. O'Neil.
\newblock Generalized isolation level definitions.
\newblock In {\em Proceedings of the 16th International Conference on Data
  Engineering, San Diego, California, USA, February 28 - March 3, 2000}, pages
  67--78, 2000.
\newblock URL: \url{https://doi.org/10.1109/ICDE.2000.839388}, \href
  {http://dx.doi.org/10.1109/ICDE.2000.839388}
  {\path{doi:10.1109/ICDE.2000.839388}}.

\bibitem{BA16}
Peter Bailis, Alan Fekete, Ali Ghodsi, Joseph~M. Hellerstein, and Ion Stoica.
\newblock Scalable atomic visibility with {RAMP} transactions.
\newblock {\em {ACM} Trans. Database Syst.}, 41(3):15:1--15:45, 2016.
\newblock URL: \url{http://doi.acm.org/10.1145/2909870}, \href
  {http://dx.doi.org/10.1145/2909870} {\path{doi:10.1145/2909870}}.

\bibitem{BE95}
Hal Berenson, Philip~A. Bernstein, Jim Gray, Jim Melton, Elizabeth~J. O'Neil,
  and Patrick~E. O'Neil.
\newblock A critique of {ANSI} {SQL} isolation levels.
\newblock In {\em Proceedings of the 1995 {ACM} {SIGMOD} International
  Conference on Management of Data, San Jose, California, May 22-25, 1995.},
  pages 1--10, 1995.
\newblock URL: \url{http://doi.acm.org/10.1145/223784.223785}, \href
  {http://dx.doi.org/10.1145/223784.223785} {\path{doi:10.1145/223784.223785}}.

\bibitem{BE16}
Giovanni Bernardi and Alexey Gotsman.
\newblock Robustness against consistency models with atomic visibility.
\newblock In {\em 27th International Conference on Concurrency Theory, {CONCUR}
  2016, August 23-26, 2016, Qu{\'{e}}bec City, Canada}, pages 7:1--7:15, 2016.
\newblock URL: \url{https://doi.org/10.4230/LIPIcs.CONCUR.2016.7}, \href
  {http://dx.doi.org/10.4230/LIPIcs.CONCUR.2016.7}
  {\path{doi:10.4230/LIPIcs.CONCUR.2016.7}}.

\bibitem{BE87}
Philip~A. Bernstein, Vassco Hadzilacos, and Nathan Goodman.
\newblock {\em Concurrency Control and Recovery in Database Systems}.
\newblock Addison-Wesley Longman Publishing Co., Inc., Boston, MA, USA, 1987.

\bibitem{BR17}
Lucas Brutschy, Dimitar Dimitrov, Peter M{\"{u}}ller, and Martin~T. Vechev.
\newblock Serializability for eventual consistency: criterion, analysis, and
  applications.
\newblock In {\em Proceedings of the 44th {ACM} {SIGPLAN} Symposium on
  Principles of Programming Languages, {POPL} 2017, Paris, France, January
  18-20, 2017}, pages 458--472, 2017.
\newblock URL: \url{http://dl.acm.org/citation.cfm?id=3009895}.

\bibitem{BR18}
Lucas Brutschy, Dimitar Dimitrov, Peter M{\"{u}}ller, and Martin~T. Vechev.
\newblock Static serializability analysis for causal consistency.
\newblock In {\em Proceedings of the 39th {ACM} {SIGPLAN} Conference on
  Programming Language Design and Implementation, {PLDI} 2018, Philadelphia,
  PA, USA, June 18-22, 2018}, pages 90--104, 2018.
\newblock URL: \url{http://doi.acm.org/10.1145/3192366.3192415}, \href
  {http://dx.doi.org/10.1145/3192366.3192415}
  {\path{doi:10.1145/3192366.3192415}}.

\bibitem{BU12}
Sebastian Burckhardt, Daan Leijen, Manuel F{\"{a}}hndrich, and Mooly Sagiv.
\newblock Eventually consistent transactions.
\newblock In {\em Programming Languages and Systems - 21st European Symposium
  on Programming, {ESOP} 2012, Held as Part of the European Joint Conferences
  on Theory and Practice of Software, {ETAPS} 2012, Tallinn, Estonia, March 24
  - April 1, 2012. Proceedings}, pages 67--86, 2012.
\newblock URL: \url{https://doi.org/10.1007/978-3-642-28869-2_4}, \href
  {http://dx.doi.org/10.1007/978-3-642-28869-2_4}
  {\path{doi:10.1007/978-3-642-28869-2_4}}.

\bibitem{BU15}
Sebastian Burckhardt, Daan Leijen, Jonathan Protzenko, and Manuel
  F{\"{a}}hndrich.
\newblock Global sequence protocol: {A} robust abstraction for replicated
  shared state.
\newblock In {\em 29th European Conference on Object-Oriented Programming,
  {ECOOP} 2015, July 5-10, 2015, Prague, Czech Republic}, pages 568--590, 2015.
\newblock URL: \url{https://doi.org/10.4230/LIPIcs.ECOOP.2015.568}, \href
  {http://dx.doi.org/10.4230/LIPIcs.ECOOP.2015.568}
  {\path{doi:10.4230/LIPIcs.ECOOP.2015.568}}.

\bibitem{CA09}
Michael~J. Cahill, Uwe R{\"{o}}hm, and Alan~David Fekete.
\newblock Serializable isolation for snapshot databases.
\newblock {\em {ACM} Trans. Database Syst.}, 34(4):20:1--20:42, 2009.
\newblock URL: \url{http://doi.acm.org/10.1145/1620585.1620587}, \href
  {http://dx.doi.org/10.1145/1620585.1620587}
  {\path{doi:10.1145/1620585.1620587}}.

\bibitem{CE15}
Andrea Cerone, Giovanni Bernardi, and Alexey Gotsman.
\newblock A framework for transactional consistency models with atomic
  visibility.
\newblock In {\em 26th International Conference on Concurrency Theory, {CONCUR}
  2015, Madrid, Spain, September 1.4, 2015}, pages 58--71, 2015.
\newblock URL: \url{https://doi.org/10.4230/LIPIcs.CONCUR.2015.58}, \href
  {http://dx.doi.org/10.4230/LIPIcs.CONCUR.2015.58}
  {\path{doi:10.4230/LIPIcs.CONCUR.2015.58}}.

\bibitem{CE16}
Andrea Cerone and Alexey Gotsman.
\newblock Analysing snapshot isolation.
\newblock In {\em Proceedings of the 2016 {ACM} Symposium on Principles of
  Distributed Computing, {PODC} 2016, Chicago, IL, USA, July 25-28, 2016},
  pages 55--64, 2016.
\newblock URL: \url{http://doi.acm.org/10.1145/2933057.2933096}, \href
  {http://dx.doi.org/10.1145/2933057.2933096}
  {\path{doi:10.1145/2933057.2933096}}.

\bibitem{CE17}
Andrea Cerone, Alexey Gotsman, and Hongseok Yang.
\newblock Algebraic laws for weak consistency.
\newblock In {\em 28th International Conference on Concurrency Theory, {CONCUR}
  2017, September 5-8, 2017, Berlin, Germany}, pages 26:1--26:18, 2017.
\newblock URL: \url{https://doi.org/10.4230/LIPIcs.CONCUR.2017.26}, \href
  {http://dx.doi.org/10.4230/LIPIcs.CONCUR.2017.26}
  {\path{doi:10.4230/LIPIcs.CONCUR.2017.26}}.

\bibitem{CR17}
Natacha Crooks, Youer Pu, Lorenzo Alvisi, and Allen Clement.
\newblock Seeing is believing: {A} client-centric specification of database
  isolation.
\newblock In {\em Proceedings of the {ACM} Symposium on Principles of
  Distributed Computing, {PODC} 2017, Washington, DC, USA, July 25-27, 2017},
  pages 73--82, 2017.
\newblock URL: \url{http://doi.acm.org/10.1145/3087801.3087802}, \href
  {http://dx.doi.org/10.1145/3087801.3087802}
  {\path{doi:10.1145/3087801.3087802}}.

\bibitem{FE05b}
Alan Fekete.
\newblock Allocating isolation levels to transactions.
\newblock In {\em Proceedings of the Twenty-fourth {ACM} {SIGACT-SIGMOD-SIGART}
  Symposium on Principles of Database Systems, June 13-15, 2005, Baltimore,
  Maryland, {USA}}, pages 206--215, 2005.
\newblock URL: \url{http://doi.acm.org/10.1145/1065167.1065193}, \href
  {http://dx.doi.org/10.1145/1065167.1065193}
  {\path{doi:10.1145/1065167.1065193}}.

\bibitem{FE05a}
Alan Fekete, Dimitrios Liarokapis, Elizabeth~J. O'Neil, and Patrick E.~O'Neil
  a~fnd Dennis E.~Shasha.
\newblock Making snapshot isolation serializable.
\newblock {\em {ACM} Trans. Database Syst.}, 30(2):492--528, 2005.
\newblock URL: \url{http://doi.acm.org/10.1145/1071610.1071615}, \href
  {http://dx.doi.org/10.1145/1071610.1071615}
  {\path{doi:10.1145/1071610.1071615}}.

\bibitem{GI02}
Seth Gilbert and Nancy~A. Lynch.
\newblock Brewer's conjecture and the feasibility of consistent, available,
  partition-tolerant web services.
\newblock {\em {SIGACT} News}, 33(2):51--59, 2002.
\newblock URL: \url{http://doi.acm.org/10.1145/564585.564601}, \href
  {http://dx.doi.org/10.1145/564585.564601} {\path{doi:10.1145/564585.564601}}.

\bibitem{GO16}
Alexey Gotsman, Hongseok Yang, Carla Ferreira, Mahsa Najafzadeh, and Marc
  Shapiro.
\newblock 'cause i'm strong enough: reasoning about consistency choices in
  distributed systems.
\newblock In {\em Proceedings of the 43rd Annual {ACM} {SIGPLAN-SIGACT}
  Symposium on Principles of Programming Languages, {POPL} 2016, St.
  Petersburg, FL, USA, January 20 - 22, 2016}, pages 371--384, 2016.
\newblock URL: \url{http://doi.acm.org/10.1145/2837614.2837625}, \href
  {http://dx.doi.org/10.1145/2837614.2837625}
  {\path{doi:10.1145/2837614.2837625}}.

\bibitem{JO07}
Sudhir Jorwekar, Alan Fekete, Krithi Ramamritham, and S.~Sudarshan.
\newblock Automating the detection of snapshot isolation anomalies.
\newblock In {\em Proceedings of the 33rd International Conference on Very
  Large Data Bases, University of Vienna, Austria, September 23-27, 2007},
  pages 1263--1274, 2007.
\newblock URL:
  \url{http://www.vldb.org/conf/2007/papers/industrial/p1263-jorwekar.pdf}.

\bibitem{KA18}
Gowtham Kaki, Kartik Nagar, Mahsa Najafzadeh, and Suresh Jagannathan.
\newblock Alone together: compositional reasoning and inference for weak
  isolation.
\newblock {\em {PACMPL}}, 2({POPL}):27:1--27:34, 2018.
\newblock URL: \url{http://doi.acm.org/10.1145/3158115}, \href
  {http://dx.doi.org/10.1145/3158115} {\path{doi:10.1145/3158115}}.

\bibitem{LL11}
Wyatt Lloyd, Michael~J. Freedman, Michael Kaminsky, and David~G. Andersen.
\newblock Don't settle for eventual: scalable causal consistency for wide-area
  storage with {COPS}.
\newblock In {\em Proceedings of the 23rd {ACM} Symposium on Operating Systems
  Principles 2011, {SOSP} 2011, Cascais, Portugal, October 23-26, 2011}, pages
  401--416, 2011.
\newblock URL: \url{http://doi.acm.org/10.1145/2043556.2043593}, \href
  {http://dx.doi.org/10.1145/2043556.2043593}
  {\path{doi:10.1145/2043556.2043593}}.

\bibitem{MU08}
Madan Musuvathi.
\newblock Systematic concurrency testing using {CHESS}.
\newblock In {\em Proceedings of the 6th Workshop on Parallel and Distributed
  Systems: Testing, Analysis, and Debugging, held in conjunction with the {ACM}
  {SIGSOFT} International Symposium on Software Testing and Analysis {(ISSTA}
  2008), {PADTAD} 2008, Seattle, Washington, USA, July 20-21, 2008}, page~10,
  2008.
\newblock URL: \url{http://doi.acm.org/10.1145/1390841.1390851}, \href
  {http://dx.doi.org/10.1145/1390841.1390851}
  {\path{doi:10.1145/1390841.1390851}}.

\bibitem{SI15}
K.~C. Sivaramakrishnan, Gowtham Kaki, and Suresh Jagannathan.
\newblock Declarative programming over eventually consistent data stores.
\newblock In {\em Proceedings of the 36th {ACM} {SIGPLAN} Conference on
  Programming Language Design and Implementation, Portland, OR, USA, June
  15-17, 2015}, pages 413--424, 2015.
\newblock URL: \url{http://doi.acm.org/10.1145/2737924.2737981}, \href
  {http://dx.doi.org/10.1145/2737924.2737981}
  {\path{doi:10.1145/2737924.2737981}}.

\bibitem{SO11}
Yair Sovran, Russell Power, Marcos~K. Aguilera, and Jinyang Li.
\newblock Transactional storage for geo-replicated systems.
\newblock In {\em Proceedings of the 23rd {ACM} Symposium on Operating Systems
  Principles 2011, {SOSP} 2011, Cascais, Portugal, October 23-26, 2011}, pages
  385--400, 2011.
\newblock URL: \url{http://doi.acm.org/10.1145/2043556.2043592}, \href
  {http://dx.doi.org/10.1145/2043556.2043592}
  {\path{doi:10.1145/2043556.2043592}}.

\bibitem{TE13}
Douglas~B. Terry, Vijayan Prabhakaran, Ramakrishna Kotla, Mahesh Balakrishnan,
  Marcos~K. Aguilera, and Hussam Abu{-}Libdeh.
\newblock Consistency-based service level agreements for cloud storage.
\newblock In {\em {ACM} {SIGOPS} 24th Symposium on Operating Systems
  Principles, {SOSP} 13, Farmington, PA, USA, November 3-6, 2013}, pages
  309--324, 2013.
\newblock URL: \url{http://doi.acm.org/10.1145/2517349.2522731}, \href
  {http://dx.doi.org/10.1145/2517349.2522731}
  {\path{doi:10.1145/2517349.2522731}}.

\bibitem{WA17}
Todd Warszawski and Peter Bailis.
\newblock Acidrain: Concurrency-related attacks on database-backed web
  applications.
\newblock In {\em Proceedings of the 2017 {ACM} International Conference on
  Management of Data, {SIGMOD} Conference 2017, Chicago, IL, USA, May 14-19,
  2017}, pages 5--20, 2017.
\newblock URL: \url{http://doi.acm.org/10.1145/3035918.3064037}, \href
  {http://dx.doi.org/10.1145/3035918.3064037}
  {\path{doi:10.1145/3035918.3064037}}.

\bibitem{ZE14}
Kamal Zellag and Bettina Kemme.
\newblock Consistency anomalies in multi-tier architectures: automatic
  detection and prevention.
\newblock {\em {VLDB} J.}, 23(1):147--172, 2014.
\newblock URL: \url{https://doi.org/10.1007/s00778-013-0318-x}, \href
  {http://dx.doi.org/10.1007/s00778-013-0318-x}
  {\path{doi:10.1007/s00778-013-0318-x}}.

\bibitem{ZH13}
Yang Zhang, Russell Power, Siyuan Zhou, Yair Sovran, Marcos~K. Aguilera, and
  Jinyang Li.
\newblock Transaction chains: achieving serializability with low latency in
  geo-distributed storage systems.
\newblock In {\em {ACM} {SIGOPS} 24th Symposium on Operating Systems
  Principles, {SOSP} '13, Farmington, PA, USA, November 3-6, 2013}, pages
  276--291, 2013.
\newblock URL: \url{http://doi.acm.org/10.1145/2517349.2522729}, \href
  {http://dx.doi.org/10.1145/2517349.2522729}
  {\path{doi:10.1145/2517349.2522729}}.

\end{thebibliography}

\appendix
\section{Complete FOL Encoding}

\begin{algorithm}
\caption{Algorithm to extract conditions for SQL statements in Transactional program $\mathcal{T}$, initially called with $\varphi = true$}
\begin{algorithmic}[1]
\Procedure{ExtractConds}{\C{c}, $\varphi$}
  \Match{$\C{c}$}
    \Case{$\C{c} \in \C{Stmts}(\mathcal{T})$ }
      \State $\Lambda_{\mathcal{T}}(\C{c}) \leftarrow \varphi$
    \EndCase
    \Case{$\C{c}_1\ ;\ \C{c}_2$}
      \State ExtractConds($\C{c}_1$, $\varphi$)
      \State ExtractConds($\C{c}_2$, $\varphi$)
    \EndCase
    \Case{\C{IF} $\phi$ \C{THEN} \C{c}$_1$ \C{ELSE} \C{c}$_2$}
      \State ExtractConds($\C{c}_1$, $\varphi \wedge \phi$)
      \State ExtractConds($\C{c}_2$, $\varphi \wedge \neg \phi$)
    \EndCase
    \Case{\C{FOREACH} \C{v}$_1$ \C{IN} \C{v}$_2$ \C{DO} \C{c$_1$} \C{END}}
      \State ExtractConds($\C{c}$, $\varphi$)
    \EndCase
  \EndMatch
\EndProcedure
\end{algorithmic}   
\end{algorithm}

The procedure \rulelabel{ExtractConds} takes the code of a transactional program $\mathcal{T}$ and populates a mapping $\Lambda_{\M{T}}$ from each SQL statement in \C{Stmts}($\mathcal{T}$) to formulaes in enclosing $\C{IF}$ conditionals. The procedure is called with the code of all transactional programs with the initial predicate $\varphi = \F{true}$.

\begin{figure}[h]
\begin{mathpar}
\small
\begin{array}{lclr}
\llbracket \phi_1 \circ \phi_2 \rrbracket_{t} & = & (\varphi_1 \wedge \varphi_2, \psi_1 \circ  \psi_2) & \llbracket \phi_1 \rrbracket_{t} = (\varphi_1, \psi_1)\\
& & &  \llbracket \phi_2 \rrbracket_{t} = (\varphi_2, \psi_2)\\
\llbracket \neg \phi \rrbracket_{t} & = & (\varphi, \neg \psi) & \llbracket \phi \rrbracket_{t} = (\varphi, \psi)\\
\llbracket \C{v} \odot e \rrbracket_{t} & = & (\varphi_1 \wedge \varphi_2, \psi_1 \odot \psi_2) & \llbracket v \rrbracket_{t} = (\varphi_1, \psi_1)\\
& & & \llbracket e \rrbracket_{t} = (\varphi_2, \psi_2)\\
\llbracket \C{v} = \C{NULL} \rrbracket_{t} & = & (\exists (r_1,\ldots, r_{\mathcal{D}(\C{v})} : R).\ \bigwedge_{i=1}^{\mathcal{D}(\C{v})} \mathcal{V} (\llbracket r_i \in \C{LVar}(\C{v},i) \rrbracket_{t}), &  \F{fresh}(r_1, \ldots, r_{\mathcal{D}(\C{v})}, r)\\
 & & \forall (r : R).\neg \rho_{\C{v}}(r_1, \ldots, r_{\mathcal{D}(\C{v})}, r, t))\\
\llbracket r \in \C{v} \rrbracket_{t} & = & (\exists (r_1,r_2,\ldots, r_{\mathcal{D}(\C{v})} : R).\  \bigwedge_{i=1}^{\mathcal{D}(\C{v})} \mathcal{V} (\llbracket r_i \in \C{LVar}(\C{v},i) \rrbracket_{t}), & \F{fresh}(r_1, \ldots, r_{\mathcal{D}(\C{v})})\\
 & & \rho_{\C{v}}(r_1, \ldots, r_{\mathcal{D}(\C{v})}, r, t)) \\
\llbracket \C{v}_1 \in \C{v}_2 \rrbracket_{t} & = & (\varphi_1 \wedge \varphi_2, \psi_2) &  \llbracket \C{v}_1 \rrbracket_{t} = (\varphi_1, \psi_1)\\
& & & \llbracket \psi_1 \in \C{v}_2 \rrbracket_{t} = (\varphi_2, \psi_2)\\
\llbracket e_1 \oplus e_2 \rrbracket_{t} & = & (\varphi_1 \wedge \varphi_2, \psi_1 \oplus \psi_2) &  \llbracket e_1 \rrbracket_{t} = (\varphi_1, \psi_1)\\
& & &  \llbracket e_2 \rrbracket_{t} = (\varphi_2, \psi_2)\\
\llbracket \C{v} \rrbracket_{t} & = &  (\exists (r_1,r_2,\ldots, r_{\mathcal{D}(\C{v})} : R).\  \bigwedge_{i=1}^{\mathcal{D}(\C{v})} \mathcal{V}(\llbracket r_i \in \C{LVar}(\C{v},i) \rrbracket_{t}), & \F{fresh}(r_1, \ldots, r_{\mathcal{D}(\C{v})})\\
& & \rho_{\C{v}}(r_1, \ldots, r_{\mathcal{D}(\C{v})}, t))\\
\llbracket n \rrbracket_{t} & = &  (\F{true},n)\\
\end{array}
\end{mathpar}
\caption{Encoding conditionals and \C{WHERE} clauses}
\label{Fig:conds-app}
\vspace{-10pt}
\end{figure}

\begin{figure}[h]
\begin{mathpar}
\begin{array}{lclr}
\llbracket \phi_1 \circ \phi_2 \rrbracket_{t,r} & = & (\varphi_1 \wedge \varphi_2, \psi_1 \circ \psi_2) & \llbracket \phi_1 \rrbracket_{t,r} = (\varphi_1, \psi_1)\\
& & & \llbracket \phi_2 \rrbracket_{t,r} = (\varphi_2, \psi_2)\\
\llbracket \neg \phi \rrbracket_{t,r} & = & (\varphi, \neg \psi) & \llbracket \phi \rrbracket_{t,r} = (\varphi, \psi)\\
\llbracket \C{f} \odot e \rrbracket_{t,r} & = & \begin{cases} (\varphi, \rho_{\C{f}}(r) \odot \psi) & \text{if } \C{f} \cup \mathcal{F}(e) \subseteq \C{PK} \cup \C{RO}   \\ (\F{true}, \F{true}) & \text{otherwise} \end{cases} \\
& & & \llbracket e \rrbracket_{t,r} = (\varphi, \psi)\\
\llbracket e_1 \oplus e_2 \rrbracket_{t,r} & = & (\varphi_1 \wedge \varphi_2, \psi_1 \oplus \psi_2) & \llbracket e_1 \rrbracket_{t,r} = (\varphi_1, \psi_1)\\
& & & \llbracket e_2 \rrbracket_{t,r} = (\varphi_2, \psi_2)\\
\llbracket \C{v} \rrbracket_{t,r} & = &  \llbracket \C{v} \rrbracket_{t}\\
\llbracket \C{f} \rrbracket_{t,r} & = & (\F{true}, \rho_{\C{f}}(r))\\
\llbracket n \rrbracket_{t,r} & = &  (\F{true}, n)\\
\end{array}
\end{mathpar}
\caption{Encoding \C{WHERE} clauses}
\label{Fig:where-app}
\end{figure}

Figures \ref{Fig:conds-app} and \ref{Fig:where-app} show the complete encoding of conditionals and \C{WHERE} clauses. Note that the field projection function is only used for primary key fields which are accessed within \C{WHERE} clauses. The other fields can be modified and hence the values contained within those fields are a function of both the record and the transaction instance, which we do not model. This does not affect the soundness of the approach, and since majority of queries in database transactions only use the primary key fields, the effect on precision is minimal. As an optimization, we also find read-only fields (\C{RO}) which are never modified within any transaction, and treat fields in \C{RO} in the same manner as fields in \C{PK}.

Below, we show how several known weak consistency specifications can be encoded in our vocabulary:


\begin{itemize}
\item Full Serializability : 
\begin{equation}
\forall (t_1, t_2 : \tau). \F{vis}(t_1, t_2) \Leftrightarrow \F{ar}(t_1, t_2)
\end{equation}
\item Selective Serializability for transactional programs $T_1, T_2$ : 
\begin{eqnarray}{\nonumber}
\forall (t_1, t_2 : \tau). ((\Gamma(t_1) =T_1 \wedge \Gamma(t_2) = T_2) \vee (\Gamma(t_1) = T_2 \wedge \Gamma(t_2) = T_1)) \\
\wedge \F{ar}(t_1,t_2) \Rightarrow \F{vis}(t_1,t_2)
\end{eqnarray}
\item Causal Consistency : 
\begin{equation}
\forall (t_1, t_2, t_3 : \tau). \F{vis}(t_1, t_2) \wedge \F{vis}(t_2, t_3) \Rightarrow \F{vis}(t_1, t_3)
\end{equation}
\item Prefix Consistency (equivalent to Repeatable Read in centralized databases) : 
\begin{equation}
\forall (t_1, t_2, t_3 : \tau). \F{ar}(t_1, t_2) \wedge \F{vis}(t_2, t_3) \Rightarrow \F{vis}(t_1, t_3)
\end{equation}
\item Parallel Snapshot Isolation : 
\begin{equation}
 \forall (t_1, t_2 : \tau). \F{WW}(t_1, t_2) \Rightarrow \F{vis}(t_1, t_2)
\end{equation}
\end{itemize}

We now present all the rules to compute $\eta_{\C{c}_1, \C{c}_2}^{\M{R}\rightarrow,\mathcal{T}_1, \mathcal{T}_2}$:

\rulelabel{WR-Update-Select}
$$
\RULE{\C{c}_1 \equiv \tt{UPDATE\; SET}\; f = e_c\; \tt{WHERE}\; \phi_1 \\ \C{c}_2 \equiv \tt{SELECT} \; f \;\tt{AS}\; v\; \tt{WHERE}\; \phi_2 \\ \C{c}_1 \in  \C{Stmts}(\mathcal{T}_1) \quad \C{c}_2 \in  \C{Stmts}(\mathcal{T}_2) \quad \Gamma(t_1) = \mathcal{T}_1 \quad \Gamma(t_2) = \mathcal{T}_2}
{\eta_{\C{c}_1, \C{c}_2}^{\F{WR}\rightarrow,\mathcal{T}_1, \mathcal{T}_2}(t_1, t_2) =  (\exists r.\ \mathcal{V}(\llbracket \Lambda_{\mathcal{T}_1}(\C{c}_1) \rrbracket_{t_1}) \wedge \mathcal{V}(\llbracket \phi_1 \rrbracket_{t_1, r} ) \wedge  \mathcal{V}(\llbracket  \Lambda_{\mathcal{T}_1}(\C{c}_1) \rrbracket_{t_2}) \wedge \mathcal{V}(\llbracket \phi_2 \rrbracket_{t_2, r}) \wedge \\ \C{Alive}(r, t_1) \wedge  \mathcal{V}(\llbracket \neg(v = \C{NULL}) \rrbracket_{t_2} \wedge \F{WR}(r, \F{f}, t_1, t_2))}
$$

\rulelabel{RW-Update-Select}
$$
\RULE{\C{c}_1 \equiv \tt{UPDATE\; SET}\; f = e_c\; \tt{WHERE}\; \phi_1 \\ \C{c}_2 \equiv \tt{SELECT} \; f \;\tt{AS}\; v\; \tt{WHERE}\; \phi_2 \\ \C{c}_1 \in  \C{Stmts}(\mathcal{T}_1) \quad \C{c}_2 \in  \C{Stmts}(\mathcal{T}_2) \quad \Gamma(t_1) = \mathcal{T}_1 \quad \Gamma(t_2) = \mathcal{T}_2}
{\eta_{\C{c}_2, \C{c}_1}^{\F{RW}\rightarrow,\mathcal{T}_2, \mathcal{T}_1}(t_2, t_1) =  (\exists r.\ \mathcal{V}(\llbracket \Lambda_{\mathcal{T}_1}(\C{c}_1) \rrbracket_{t_1}) \wedge \mathcal{V}(\llbracket \phi_1 \rrbracket_{t_1, r} ) \wedge  \mathcal{V}(\llbracket  \Lambda_{\mathcal{T}_1}(\C{c}_1) \rrbracket_{t_2}) \wedge \mathcal{V}(\llbracket \phi_2 \rrbracket_{t_2, r}) \wedge \\ \C{Alive}(r, t_1)  \wedge \F{RW}(r, \F{f}, t_2, t_1))}
$$

\rulelabel{WR-Insert-Select}
$$
\RULE{\C{c}_1 \equiv  \tt{INSERT\; VALUES}\; \bar{f} = \bar{e} \\ \C{c}_2 \equiv \tt{SELECT} \; f \;\tt{AS}\; v\; \tt{WHERE}\; \phi_2 \\ \C{c}_1 \in  \C{Stmts}(\mathcal{T}_1) \quad \C{c}_2 \in  \C{Stmts}(\mathcal{T}_2) \quad \Gamma(t_1) = \mathcal{T}_1 \quad \Gamma(t_2) = \mathcal{T}_2 \\ \forall (\C{f} \in \C{Fields}).\ \llbracket \bar{e}(\C{f}) \rrbracket_{t_1} = (\varphi_{\C{f}}, \psi_{\C{f}}) }
{\eta_{\C{c}_1, \C{c}_2}^{\F{WR}\rightarrow,\mathcal{T}_1, \mathcal{T}_2}(t_1, t_2) =  (\exists r.\ \mathcal{V}(\llbracket \Lambda_{\mathcal{T}_1}(\C{c}_1) \rrbracket_{t_1}) \wedge \bigwedge_{\C{f} \in \C{Fields}} \varphi_{\C{f}} \wedge \bigwedge_{\C{f} \in \C{Fields}} \rho_{\C{f}}(r) = \psi_{\C{f}} \\ \wedge  \mathcal{V}(\llbracket  \Lambda_{\mathcal{T}_1}(\C{c}_1) \rrbracket_{t_2}) \wedge \mathcal{V}(\llbracket \phi_2 \rrbracket_{t_2, r}) \wedge  \C{Alive}(r, t_2) \wedge  \mathcal{V}(\llbracket \neg(v = \C{NULL}) \rrbracket_{t_2} ) \wedge \F{WR}(r, \C{Alive}, t_1, t_2))}
$$

\rulelabel{RW-Insert-Select}
$$
\RULE{\C{c}_1 \equiv  \tt{INSERT\; VALUES}\; \bar{f} = \bar{e} \\ \C{c}_2 \equiv \tt{SELECT} \; f \;\tt{AS}\; v\; \tt{WHERE}\; \phi_2 \\ \C{c}_1 \in  \C{Stmts}(\mathcal{T}_1) \quad \C{c}_2 \in  \C{Stmts}(\mathcal{T}_2) \quad \Gamma(t_1) = \mathcal{T}_1 \quad \Gamma(t_2) = \mathcal{T}_2 \\ \forall (\C{f} \in \C{Fields}).\ \llbracket \bar{e}(\C{f}) \rrbracket_{t_1} = (\varphi_{\C{f}}, \psi_{\C{f}}) }
{\eta_{\C{c}_2, \C{c}_1}^{\F{RW}\rightarrow,\mathcal{T}_2, \mathcal{T}_1}(t_2, t_1) =  (\exists r.\ \mathcal{V}(\llbracket \Lambda_{\mathcal{T}_1}(\C{c}_1) \rrbracket_{t_1}) \wedge \bigwedge_{\C{f} \in \C{Fields}} \varphi_{\C{f}} \wedge \bigwedge_{\C{f} \in \C{Fields}} \rho_{\C{f}}(r) = \psi_{\C{f}}  \\ \wedge  \mathcal{V}(\llbracket  \Lambda_{\mathcal{T}_1}(\C{c}_1) \rrbracket_{t_2}) \wedge \mathcal{V}(\llbracket \phi_2 \rrbracket_{t_2, r}) \wedge  \neg \C{Alive}(r, t_2) \wedge  \mathcal{V}(\llbracket \neg(r \in v) \rrbracket_{t_2} ) \wedge \F{RW}(r, \C{Alive}, t_2, t_1))}
$$

\rulelabel{WR-Delete-Select}
$$
\RULE{\C{c}_1 \equiv   \tt{DELETE\; WHERE}\; \phi_1 \\ \C{c}_2 \equiv \tt{SELECT} \; f \;\tt{AS}\; v\; \tt{WHERE}\; \phi_2 \\ \C{c}_1 \in  \C{Stmts}(\mathcal{T}_1) \quad \C{c}_2 \in  \C{Stmts}(\mathcal{T}_2) \quad \Gamma(t_1) = \mathcal{T}_1 \quad \Gamma(t_2) = \mathcal{T}_2 }
{\eta_{\C{c}_1, \C{c}_2}^{\F{WR}\rightarrow,\mathcal{T}_1, \mathcal{T}_2}(t_1, t_2) =  (\exists r.\ \mathcal{V}(\llbracket \Lambda_{\mathcal{T}_1}(\C{c}_1) \rrbracket_{t_1}) \wedge \mathcal{V}(\llbracket \phi_1 \rrbracket_{t_1, r} )  \wedge  \mathcal{V}(\llbracket  \Lambda_{\mathcal{T}_1}(\C{c}_1) \rrbracket_{t_2}) \wedge \mathcal{V}(\llbracket \phi_2 \rrbracket_{t_2, r}) \wedge \\ \C{Alive}(r, t_1) \wedge  \neg \C{Alive}(r, t_2) \wedge  \mathcal{V}(\llbracket \neg(r \in v) \rrbracket_{t_2} ) \wedge \F{WR}(r, \C{Alive}, t_1, t_2))}
$$

\rulelabel{RW-Delete-Select}
$$
\RULE{\C{c}_1 \equiv   \tt{DELETE\; WHERE}\; \phi_1 \\ \C{c}_2 \equiv \tt{SELECT} \; f \;\tt{AS}\; v\; \tt{WHERE}\; \phi_2 \\ \C{c}_1 \in  \C{Stmts}(\mathcal{T}_1) \quad \C{c}_2 \in  \C{Stmts}(\mathcal{T}_2) \quad \Gamma(t_1) = \mathcal{T}_1 \quad \Gamma(t_2) = \mathcal{T}_2 }
{\eta_{\C{c}_2, \C{c}_1}^{\F{RW}\rightarrow,\mathcal{T}_2, \mathcal{T}_1}(t_2, t_1) =  (\exists r.\ \mathcal{V}(\llbracket \Lambda_{\mathcal{T}_1}(\C{c}_1) \rrbracket_{t_1}) \wedge \mathcal{V}(\llbracket \phi_1 \rrbracket_{t_1, r} )  \wedge  \mathcal{V}(\llbracket  \Lambda_{\mathcal{T}_1}(\C{c}_1) \rrbracket_{t_2}) \wedge \mathcal{V}(\llbracket \phi_2 \rrbracket_{t_2, r}) \wedge \\ \C{Alive}(r, t_1) \wedge \mathcal{V}(\llbracket \neg(v = \C{NULL}) \rrbracket_{t_2} ) \wedge \F{RW}(r, \C{Alive}, t_2, t_1))}
$$

\rulelabel{WR-Insert-Select-Max}
$$
\RULE{\C{c}_1 \equiv  \tt{INSERT\; VALUES}\; \bar{f} = \bar{e} \\ \C{c}_2 \equiv \tt{SELECT\ MAX} \; \F{f_2} \;\tt{AS}\; \C{v}\; \tt{WHERE}\; \phi_2 \\  \C{c}_1 \in  \C{Stmts}(\mathcal{T}_1) \quad \C{c}_2 \in  \C{Stmts}(\mathcal{T}_2) \quad \Gamma(t_1) = \mathcal{T}_1 \quad \Gamma(t_2) = \mathcal{T}_2\\ \llbracket \C{v} \rrbracket_{t_2} = (\varphi, \psi) \quad \forall (\C{f} \in \C{Fields}).\ \llbracket \bar{e}(\C{f}) \rrbracket_{t_1} = (\varphi_{\C{f}}, \psi_{\C{f}})}
{\eta_{\C{c}_1, \C{c}_2}^{\F{WR}\rightarrow,\mathcal{T}_1, \mathcal{T}_2}(t_1, t_2) = (\exists r.\ \mathcal{V}(\llbracket \Lambda_{\mathcal{T}_1}(\C{c}_1) \rrbracket_{t_1}) \wedge \bigwedge_{\C{f} \in \C{Fields}} \varphi_{\C{f}} \wedge \bigwedge_{\C{f} \in \C{Fields}} \rho_{\C{f}}(r) = \psi_{\C{f}} \\ \wedge  \mathcal{V}(\llbracket \Lambda_{\mathcal{T}_2}(\C{c}_2) \rrbracket_{t_2}) \wedge \mathcal{V}(\llbracket \phi_2 \rrbracket_{t_2, r}) \wedge  \C{Alive}(r, t_2) \wedge  \varphi \wedge \psi_{\C{f}_2} \leq \psi )}
$$

\rulelabel{RW-Insert-Select-Max}
$$
\RULE{\C{c}_1 \equiv  \tt{INSERT\; VALUES}\; \bar{f} = \bar{e} \\ \C{c}_2 \equiv \tt{SELECT\ MAX} \; \F{f_2} \;\tt{AS}\; \C{v}\; \tt{WHERE}\; \phi_2   \\ \C{c}_1 \in  \C{Stmts}(\mathcal{T}_1) \quad \C{c}_2 \in  \C{Stmts}(\mathcal{T}_2) \quad \Gamma(t_1) = \mathcal{T}_1 \quad \Gamma(t_2) = \mathcal{T}_2\\ \llbracket \C{v} \rrbracket_{t_2} = (\varphi, \psi) \quad \forall (\C{f} \in \C{Fields}).\ \llbracket \bar{e}(\C{f}) \rrbracket_{t_1} = (\varphi_{\C{f}}, \psi_{\C{f}})}
{\eta_{\C{c}_2, \C{c}_1}^{\F{RW}\rightarrow,\mathcal{T}_2, \mathcal{T}_1}(t_2, t_1) = (\exists r.\ \mathcal{V}(\llbracket \Lambda_{\mathcal{T}_1}(\C{c}_1) \rrbracket_{t_1}) \wedge \bigwedge_{\C{f} \in \C{Fields}} \varphi_{\C{f}} \wedge \bigwedge_{\C{f} \in \C{Fields}} \rho_{\C{f}}(r) = \psi_{\C{f}} \\ \wedge  \mathcal{V}(\llbracket \Lambda_{\mathcal{T}_2}(\C{c}_2) \rrbracket_{t_2}) \wedge \mathcal{V}(\llbracket \phi_2 \rrbracket_{t_2, r}) \wedge \neg \C{Alive}(r, t_2) \wedge \varphi \wedge \psi_{\C{f}_2} > \psi )}
$$

\rulelabel{WR-Insert-Select-Min}
$$
\RULE{\C{c}_1 \equiv  \tt{INSERT\; VALUES}\; \bar{f} = \bar{e} \\ \C{c}_2 \equiv \tt{SELECT\ MIN} \; \F{f_2} \;\tt{AS}\; \C{v}\; \tt{WHERE}\; \phi_2 \\  \C{c}_1 \in  \C{Stmts}(\mathcal{T}_1) \quad \C{c}_2 \in  \C{Stmts}(\mathcal{T}_2) \quad \Gamma(t_1) = \mathcal{T}_1 \quad \Gamma(t_2) = \mathcal{T}_2\\ \llbracket \C{v} \rrbracket_{t_2} = (\varphi, \psi) \quad \forall (\C{f} \in \C{Fields}).\ \llbracket \bar{e}(\C{f}) \rrbracket_{t_1} = (\varphi_{\C{f}}, \psi_{\C{f}})}
{\eta_{\C{c}_1, \C{c}_2}^{\F{WR}\rightarrow,\mathcal{T}_1, \mathcal{T}_2}(t_1, t_2) = (\exists r.\ \mathcal{V}(\llbracket \Lambda_{\mathcal{T}_1}(\C{c}_1) \rrbracket_{t_1}) \wedge \bigwedge_{\C{f} \in \C{Fields}} \varphi_{\C{f}} \wedge \bigwedge_{\C{f} \in \C{Fields}} \rho_{\C{f}}(r) = \psi_{\C{f}} \\ \wedge  \mathcal{V}(\llbracket \Lambda_{\mathcal{T}_2}(\C{c}_2) \rrbracket_{t_2}) \wedge \mathcal{V}(\llbracket \phi_2 \rrbracket_{t_2, r}) \wedge  \C{Alive}(r, t_2) \wedge  \varphi \wedge \psi_{\C{f}_2} \geq \psi )}
$$

\rulelabel{RW-Insert-Select-Min}
$$
\RULE{\C{c}_1 \equiv  \tt{INSERT\; VALUES}\; \bar{f} = \bar{e} \\ \C{c}_2 \equiv \tt{SELECT\ MIN} \; \F{f_2} \;\tt{AS}\; \C{v}\; \tt{WHERE}\; \phi_2  \\ \C{c}_1 \in  \C{Stmts}(\mathcal{T}_1) \quad \C{c}_2 \in  \C{Stmts}(\mathcal{T}_2) \quad \Gamma(t_1) = \mathcal{T}_1 \quad \Gamma(t_2) = \mathcal{T}_2\\ \llbracket \C{v} \rrbracket_{t_2} = (\varphi, \psi) \quad \forall (\C{f} \in \C{Fields}).\ \llbracket \bar{e}(\C{f}) \rrbracket_{t_1} = (\varphi_{\C{f}}, \psi_{\C{f}})}
{\eta_{\C{c}_2, \C{c}_1}^{\F{RW}\rightarrow,\mathcal{T}_2, \mathcal{T}_1}(t_2, t_1) = (\exists r.\ \mathcal{V}(\llbracket \Lambda_{\mathcal{T}_1}(\C{c}_1) \rrbracket_{t_1}) \wedge \bigwedge_{\C{f} \in \C{Fields}} \varphi_{\C{f}} \wedge \bigwedge_{\C{f} \in \C{Fields}} \rho_{\C{f}}(r) = \psi_{\C{f}} \\ \wedge  \mathcal{V}(\llbracket \Lambda_{\mathcal{T}_2}(\C{c}_2) \rrbracket_{t_2}) \wedge \mathcal{V}(\llbracket \phi_2 \rrbracket_{t_2, r}) \wedge  \neg \C{Alive}(r, t_2) \wedge \varphi \wedge \psi_{\C{f}_2} < \psi )}
$$

\rulelabel{WR-Delete-Select-Min}
$$
\RULE{\C{c}_1 \equiv   \tt{DELETE\; WHERE}\; \phi_1 \\ \C{c}_2 \equiv \tt{SELECT MIN} \; f \;\tt{AS}\; v\; \tt{WHERE}\; \phi_2 \\ \C{c}_1 \in  \C{Stmts}(\mathcal{T}_1) \quad \C{c}_2 \in  \C{Stmts}(\mathcal{T}_2) \quad \Gamma(t_1) = \mathcal{T}_1 \quad \Gamma(t_2) = \mathcal{T}_2 \quad \llbracket v \rrbracket_{t_2} = (\varphi, \psi) }
{\eta_{\C{c}_1, \C{c}_2}^{\F{WR}\rightarrow,\mathcal{T}_1, \mathcal{T}_2}(t_1, t_2) =  (\exists r.\ \mathcal{V}(\llbracket \Lambda_{\mathcal{T}_1}(\C{c}_1) \rrbracket_{t_1}) \wedge \mathcal{V}(\llbracket \phi_1 \rrbracket_{t_1, r} )  \wedge  \mathcal{V}(\llbracket  \Lambda_{\mathcal{T}_1}(\C{c}_1) \rrbracket_{t_2}) \wedge \mathcal{V}(\llbracket \phi_2 \rrbracket_{t_2, r}) \wedge \\ \C{Alive}(r, t_1) \wedge  \neg \C{Alive}(r, t_2) \wedge \varphi \wedge \rho_{\C{f}}(r) \geq \psi \wedge \F{WR}(r, \C{Alive}, t_1, t_2))}
$$

\rulelabel{RW-Delete-Select-Min}
$$
\RULE{\C{c}_1 \equiv   \tt{DELETE\; WHERE}\; \phi_1 \\ \C{c}_2 \equiv \tt{SELECT MIN} \; f \;\tt{AS}\; v\; \tt{WHERE}\; \phi_2 \\ \C{c}_1 \in  \C{Stmts}(\mathcal{T}_1) \quad \C{c}_2 \in  \C{Stmts}(\mathcal{T}_2) \quad \Gamma(t_1) = \mathcal{T}_1 \quad \Gamma(t_2) = \mathcal{T}_2 \quad \llbracket v \rrbracket_{t_2} = (\varphi, \psi) }
{\eta_{\C{c}_2, \C{c}_1}^{\F{RW}\rightarrow,\mathcal{T}_2, \mathcal{T}_1}(t_2, t_1) =  (\exists r.\ \mathcal{V}(\llbracket \Lambda_{\mathcal{T}_1}(\C{c}_1) \rrbracket_{t_1}) \wedge \mathcal{V}(\llbracket \phi_1 \rrbracket_{t_1, r} )  \wedge  \mathcal{V}(\llbracket  \Lambda_{\mathcal{T}_1}(\C{c}_1) \rrbracket_{t_2}) \wedge \mathcal{V}(\llbracket \phi_2 \rrbracket_{t_2, r}) \wedge \\ \C{Alive}(r, t_1) \wedge \varphi \wedge \rho_{\C{f}}(r) = \psi \wedge \F{RW}(r, \C{Alive}, t_2, t_1))}
$$

\rulelabel{WW-Update-Update}
$$
\RULE{\C{c}_1 \equiv \tt{UPDATE\; SET}\; f = e_1\; \tt{WHERE}\; \phi_1 \\ \C{c}_2 \equiv \tt{UPDATE\; SET}\; f = e_2\; \tt{WHERE}\; \phi_2 \\ \C{c}_1 \in  \C{Stmts}(\mathcal{T}_1) \quad \C{c}_2 \in  \C{Stmts}(\mathcal{T}_2) \quad \Gamma(t_1) = \mathcal{T}_1 \quad \Gamma(t_2) = \mathcal{T}_2}
{\eta_{\C{c}_1, \C{c}_2}^{\F{WW}\rightarrow,\mathcal{T}_1, \mathcal{T}_2}(t_1, t_2) =  (\exists r.\ \mathcal{V}(\llbracket \Lambda_{\mathcal{T}_1}(\C{c}_1) \rrbracket_{t_1}) \wedge \mathcal{V}(\llbracket \phi_1 \rrbracket_{t_1, r} ) \wedge  \mathcal{V}(\llbracket  \Lambda_{\mathcal{T}_1}(\C{c}_1) \rrbracket_{t_2}) \wedge \mathcal{V}(\llbracket \phi_2 \rrbracket_{t_2, r}) \wedge \\ \C{Alive}(r, t_1) \wedge  \C{Alive}(r, t_2) \wedge \F{WW}(r, \F{f}, t_1, t_2))}
$$

We now present the rules for $\eta_{\C{c}_1, \C{c}_2}^{\rightarrow \M{R},\mathcal{T}_1, \mathcal{T}_2}$:

\rulelabel{INSERT-SELECT-WR}
$$
\RULE{\C{c}_1 \equiv  \tt{INSERT\; VALUES}\; \bar{f} = \bar{e} \\ \C{c}_2 \equiv \tt{SELECT} \; f \;\tt{AS}\; v\; \tt{WHERE}\; \phi_2 \\ \C{c}_1 \in  \C{Stmts}(\mathcal{T}_1) \quad \C{c}_2 \in  \C{Stmts}(\mathcal{T}_2) \quad \Gamma(t_1) = \mathcal{T}_1 \quad \Gamma(t_2) = \mathcal{T}_2 \\ \forall (\C{f} \in \C{Fields}).\ \llbracket \bar{e}(\C{f}) \rrbracket_{t_1} = (\varphi_{\C{f}}, \psi_{\C{f}}) }
{\eta_{\C{c}_1, \C{c}_2}^{\rightarrow\F{WR},\mathcal{T}_1, \mathcal{T}_2}(t_1, t_2) =  (\exists r.\ \mathcal{V}(\llbracket \Lambda_{\mathcal{T}_1}(\C{c}_1) \rrbracket_{t_1}) \wedge \bigwedge_{\C{f} \in \C{Fields}} \varphi_{\C{f}} \wedge \bigwedge_{\C{f} \in \C{Fields}} \rho_{\C{f}}(r) = \psi_{\C{f}} \\ \wedge  \mathcal{V}(\llbracket  \Lambda_{\mathcal{T}_1}(\C{c}_1) \rrbracket_{t_2}) \wedge \mathcal{V}(\llbracket \phi_2 \rrbracket_{t_2, r}) \wedge \M{V}(\llbracket r \in v \rrbracket_{t_2})\wedge \{\C{f}\} \cup \M{F}(\phi_2) = \C{PK})}
$$

\rulelabel{INSERT-SELECT-MIN-WR}
$$
\RULE{\C{c}_1 \equiv  \tt{INSERT\; VALUES}\; \bar{f} = \bar{e} \\ \C{c}_2 \equiv \tt{SELECT} \; f \;\tt{AS}\; v\; \tt{WHERE}\; \phi_2 \\ \C{c}_1 \in  \C{Stmts}(\mathcal{T}_1) \quad \C{c}_2 \in  \C{Stmts}(\mathcal{T}_2) \quad \Gamma(t_1) = \mathcal{T}_1 \quad \Gamma(t_2) = \mathcal{T}_2 \\ \forall (\C{f} \in \C{Fields}).\ \llbracket \bar{e}(\C{f}) \rrbracket_{t_1} = (\varphi_{\C{f}}, \psi_{\C{f}}) }
{\eta_{\C{c}_1, \C{c}_2}^{\rightarrow\F{WR},\mathcal{T}_1, \mathcal{T}_2}(t_1, t_2) =  (\exists r.\ \mathcal{V}(\llbracket \Lambda_{\mathcal{T}_1}(\C{c}_1) \rrbracket_{t_1}) \wedge \bigwedge_{\C{f} \in \C{Fields}} \varphi_{\C{f}} \wedge \bigwedge_{\C{f} \in \C{Fields}} \rho_{\C{f}}(r) = \psi_{\C{f}} \\ \wedge  \mathcal{V}(\llbracket  \Lambda_{\mathcal{T}_1}(\C{c}_1) \rrbracket_{t_2}) \wedge \mathcal{V}(\llbracket \phi_2 \rrbracket_{t_2, r}) \wedge \M{V}(\llbracket r \in v \rrbracket_{t_2})\wedge \{\C{f}\} \cup \M{F}(\phi_2) = \C{PK})}
$$

The contents of the primary key fields uniquely determine a record, so that if two records have the same values in their \C{PK} fields, then they will also have same values in all other fields:
\begin{equation}
\boxed
{
\varphi_{pk} = \forall (r_1, r_2 : R).\ (\bigwedge_{\C{f} \in \C{PK}} \C{f}(r_1) = \C{f}(r_2)) \Rightarrow r_1 = r_2
}
\end{equation}

Any record in the output of a \C{SELECT} query must obey the \C{WHERE} clause of the query:
\begin{equation}
\boxed
{
\varphi_{\C{SELECT}} = \bigwedge_{\M{T} \in \mathbb{T}} \bigwedge_{\substack{\C{c} \in \C{Stmts}(\M{T}): \\ \C{c} \equiv \tt{SELECT} \; f \;\tt{AS}\; v\; \tt{WHERE}\; \phi}} \forall (t:\tau)(r:R).\ (\Gamma(t) = \M{T} \wedge \M{V}(\llbracket \Lambda_{\M{T}}(\C{c}) \rrbracket_{t}) \wedge \M{V}(\llbracket r \in v \rrbracket_{t})) \Rightarrow \llbracket \phi \rrbracket_{t,r}
}
\end{equation}

If a record is \C{Alive} to a transaction instance, then there must exist some transaction instance which inserts that record:
\begin{equation}
\boxed
{
\begin{split}
\varphi_{\C{INSERT}} = \forall (r:R).\ (\exists (t_1 : \tau).\ \C{Alive}(r,t_1)) \Rightarrow \bigvee_{\substack{\M{T} \in \mathbb{T}:\\ \tt{INSERT\; VALUES}\; \bar{f} = \bar{e} \in \C{Stmts}(\M{T}) }} \bigvee_{\substack{\C{c} \in \C{Stmts}(\M{T}): \\ \C{c} \equiv \tt{INSERT\; VALUES}\; \bar{f} = \bar{e}}} \\ (\exists (t_2 : \tau).\  \Gamma(t_2) = \M{T} \wedge \M{V}(\llbracket \Lambda_{\M{T}}(\C{c}) \rrbracket_{t_2}) \wedge \bigwedge_{\substack{\C{f} \in \C{Fields} \\  \llbracket \bar{e}(\C{f}) \rrbracket_{t_2} = (\varphi_{\C{f}}, \psi_{\C{f}})}} \varphi_{\C{f}} \wedge (\rho_{\C{f}}(r) = \psi_{\C{f}}) )
\end{split}
}
\end{equation}

\section{Full Operational Semantics}

We now describe the operational semantics of the transactional programs which results in abstract executions. This is an interleaving semantics where we can non-deterministically decide to begin a new instantiation of a transactional program. The system state is stored in terms of the committed transaction instances, the $\F{vis}$ and $\F{ar}$ relations among them, and a running pool of transaction instances. When a new execution of a transactional program begins, a subset of the committed transaction instances is non-deterministically selected to be made visible to the new instance. A view of the database is reconstructed based on the set of visible transactions and the $\F{ar}$ relation, and all queries of the newly executing transaction instance are answered on the basis of this view. The newly executing transaction instance is added to the pool of running transactions. At any point, any transaction instance from the running pool can be non-deterministically selected for execution. Any new event generated during the execution of a transaction instance is stored in the running pool but is not made visible to other transaction instances. Finally, when a transaction instance wants to commit, it is checked whether constraints of the weak consistency and weak isolation model ($\Psi$) are satisfied if the instance were to commit, and if yes then the instance is allowed to commit and is added to the set of committed transaction instances. 

Let $\Sigma$ denote the set of committed transaction instances. The state is maintained as a tuple $(\Sigma, \F{vis}, \F{ar}, R, \Gamma)$ where $R$ is the set of running transaction instances and $\Gamma : \Sigma \cup R \rightarrow \mathbb{T}$ maps transaction instances to the transactional programs which generated them. A running transaction instance is maintained as a tuple $r = (t, \varepsilon, c, \Delta, \Sigma_r, \theta)$, where $t$ is the unique instance ID, $\varepsilon$ is the set of events generated by the instance, $c$ is the program to be executed, $\Delta$ is the view of the database, $\Sigma_r$ is the set of committed transaction instances visible to $r$, and $\theta : \bar{v}_p(\Gamma(r)) \cup \bar{v}_l(\Gamma(r)) \rightarrow \mathbb{V}$ provides valuations of the parameter and local variables of the transaction instance. Note that $\bar{v}_p(\M{T})$ and $\bar{v}_l(\M{T})$ denote the set of parameter and local variables respectively of $\M{T}$. The view of the database $\Delta : (\mathbb{Z})^{|PK|} \rightarrow (\texttt{Fields} \setminus \texttt{PK}) \rightarrow \mathbb{Z}$ is defined as $\varsigma(\Sigma_r, \F{ar})$ and is constructed from the set of committed transaction instances visible to the running instance $\Sigma_r$ (which was decided when the running instance began its execution) in the following manner: 
\begin{equation}
\forall r \in \mathbb{Z}^{PK}, \Delta(r)(\C{f}) = n \Leftrightarrow \F{MAX_{\F{ar}}}([\Sigma_r]_{<\C{wri}(r,f)>}) \vdash \C{wri}(r,f,n)
\end{equation}

Below, we present all the rules of the operational semantics : 

\rulelabel{E-Spawn}
$$
\RULE{\mathcal{T} \in \mathbb{T} \quad \Sigma' \subseteq \Sigma \quad \Delta = \varsigma(\Sigma', \F{ar}) \quad \theta(\bar{v}_p(\mathcal{T})) \in \mathbb{Z} \quad t \in \C{TID} \quad \Gamma' = \Gamma \cup \{(r, \mathcal{T})\}\\ r = (t, \{\}, c(\mathcal{T}), \Delta, \Sigma', \theta)}{(\Sigma, \F{vis}, \F{ar}, R, \Gamma) \rightarrow (\Sigma, \F{vis}, \F{ar}, R \cup \{r\}, \Gamma')}
$$ 

\rulelabel{E-Step}
$$
\RULE{r \rightarrow r'}{(\Sigma, \F{vis}, \F{ar}, R \cup \{r\}, \Gamma) \rightarrow (\Sigma, \F{vis}, \F{ar}, R \cup \{r'\}, \Gamma) }
$$

\rulelabel{E-Commit}
$$
\RULE{r = (t, \varepsilon, \C{SKIP}, \Delta, \Sigma_r, \theta) \quad \sigma = (t, \varepsilon) \quad \F{vis'} = \F{vis} \cup \{(\sigma', \sigma)\ |\ \sigma' \in \Sigma_r\} \\
\F{ar'} = \F{ar} \cup \{(\sigma', \sigma)\ |\ \sigma' \in \Sigma\} \quad \Sigma' = \Sigma \cup \{\sigma\} \\ \Gamma' = \Gamma \cup \{(\sigma, \Gamma(r))\} \setminus \{(r, \Gamma(r))\} \quad \Psi(\Sigma', \F{vis'}, \F{ar'}, \Gamma') }{(\Sigma, \F{vis}, \F{ar}, R \cup \{r\}, \Gamma) \rightarrow (\Sigma', \F{vis'}, \F{ar'}, R, \Gamma') }
$$

\rulelabel{E-Select}
$$
\RULE{\varepsilon' = \varepsilon \cup \{\C{rd}(r,f',n)\ |\ r \in \mathcal{R} \wedge f' \in \mathcal{F}(\phi) \wedge \llbracket f' \rrbracket_{r, \Delta, \theta} = n \} \\ \cup \{\C{rd}(r, \bar{f}(i), n)\ |\ \llbracket \phi \rrbracket_{r, \Delta, \theta} \wedge \llbracket \bar{f}(i) \rrbracket_{r, \Delta, \theta} = n \wedge 1 \leq i \leq \C{len}(\bar{f})\}  \\ \cup \{\C{rd}(r, \C{Alive}, n)\ |\ \llbracket \phi \rrbracket_{r,\Delta, \theta} \wedge \Delta(r)(\C{Alive}) = n\} \\ s = \{\Delta(r)(\bar{f})\ |\ \llbracket \phi \rrbracket_{\Delta, \theta}(r) \wedge \Delta(r)(\C{Alive}) = 1\} \quad  \theta' = \theta[v \rightarrow s]}{(t, \varepsilon, \C{SELECT} \; \bar{f} \;\C{AS}\; v\; \C{WHERE}\; \phi, \Delta, \Sigma_r, \theta) \rightarrow (t, \varepsilon', \C{SKIP}, \Delta, \Sigma_r, \theta')}
$$

\rulelabel{E-Insert}
$$
\RULE{r \in \mathbb{Z}^{|\C{PK}|} \quad \forall \C{f} \in \C{PK}. r(\C{f}) = \llbracket \bar{e}_c(\C{f}) \rrbracket_{\theta}\\ \varepsilon' = \varepsilon \cup \{\C{wri}(r, \C{Alive}, 1)\} \cup \{\C{wri}(r, \C{f}, n) | \C{f} \in \C{Fields} \setminus \C{PK},\ \llbracket \bar{e}_c \rrbracket_{\theta}(\C{f}) = n \}}{(t,\varepsilon, \C{INSERT VALUES}\; \bar{f} = \bar{e}_c, \Delta, \Sigma_r, \theta) \rightarrow (t, \varepsilon', \C{SKIP}, \Delta, \Sigma_r, \theta)}
$$

\rulelabel{E-Update}
$$
\RULE{\varepsilon' = \varepsilon \cup \{\C{rd}(r,f',n)\ |\ f' \in \mathcal{F}(\phi) \wedge \llbracket f' \rrbracket_{r, \Delta, \theta} = n \} \\ \cup \{\C{wri}(r, f, n)\ |\ \llbracket \phi \rrbracket_{r, \Delta, \theta} \wedge \llbracket e_c \rrbracket_{\theta} = n \wedge \Delta(r)(\C{Alive}) = 1\} \cup \\ \{\C{rd}(r, \C{Alive}, n)\ |\ \llbracket \phi \rrbracket_{r, \Delta, \theta} \wedge \Delta(r)(\C{Alive}) = n\}}{(t, \varepsilon, \C{UPDATE SET}\; f = e_c\; \C{WHERE}\; \phi, \Delta, \Sigma_r, \theta) \rightarrow (t, \varepsilon', \C{SKIP}, \Delta, \Sigma_r, \theta)}
$$

\rulelabel{E-Delete}
$$
\RULE{\varepsilon' = \varepsilon \cup \{\C{rd}(r,f',n)\ |\ f' \in \mathcal{F}(\phi) \wedge \llbracket f' \rrbracket_{r, \Delta, \theta} = n \} \cup \{\C{wri}(r, \C{Alive}, 0)\ |\ \llbracket \phi \rrbracket_{r, \Delta, \theta}\}}{(t, \varepsilon, \C{DELETE WHERE}\; \phi, \Delta, \Sigma_r, \theta) \rightarrow (t, \varepsilon', \C{SKIP}, \Delta, \Sigma_r, \theta)}
$$

\rulelabel{E-Select-Count}
$$
\RULE{\varepsilon' = \varepsilon \cup \{\C{rd}(r,f',n)\ |\ f' \in \mathcal{F}(\phi) \wedge  \llbracket f' \rrbracket_{r, \Delta, \theta} = n \} \\ \cup \{\C{rd}(r, \C{Alive}, n)\ |\ \llbracket \phi \rrbracket_{r,\Delta, \theta} \wedge \Delta(r)(\C{Alive}) = n\} \\ s = |\{r\ |\ \llbracket \phi \rrbracket_{r, \Delta, \theta} \wedge \Delta(r)(\C{Alive}) = 1\}| \quad  \theta' = \theta[v \rightarrow s]}{(t, \varepsilon, \C{SELECT\ COUNT} \; f \;\C{AS}\; v\; \C{WHERE}\; \phi, \Delta, \Sigma_r, \theta) \rightarrow (t, \varepsilon', \C{SKIP}, \Delta, \Sigma_r, \theta')}
$$

\rulelabel{E-Select-Max}
$$
\RULE{s = \F{MAX}\{\llbracket f \rrbracket_{r, \Delta, \theta}\ |\ \llbracket \phi \rrbracket_{r, \Delta, \theta} \wedge \Delta(r)(\C{Alive}) = 1\} \quad  \theta' = \theta[v \rightarrow s] \\ \varepsilon' = \varepsilon \cup \{\C{rd}(r,f',n)\ |\ f' \in \mathcal{F}(\phi) \wedge \llbracket f' \rrbracket_{r, \Delta, \theta} = n \} \\ \cup \{\C{rd}(r, f, n)\ |\ \llbracket \phi \rrbracket_{r, \Delta, \theta} \wedge \llbracket f \rrbracket_{r, \Delta, \theta} = n \wedge n \geq s\} \\ \cup \{\C{rd}(r, \C{Alive}, n)\ |\ \llbracket \phi \rrbracket_{r, \Delta, \theta} \wedge \Delta(r)(\C{Alive}) = n \wedge \llbracket f \rrbracket_{r, \Delta, \theta} \geq s\} }{(t, \varepsilon, \C{SELECT} \; \C{MAX}\; f \;\C{AS}\; v\; \C{WHERE}\; \phi, \Delta, \Sigma_r, \theta) \rightarrow (t, \varepsilon', \C{SKIP}, \Delta, \Sigma_r, \theta')}
$$

\rulelabel{E-Select-Min}
$$
\RULE{ s = \F{MIN}\{\llbracket f \rrbracket_{r, \Delta, \theta}\ |\ \llbracket \phi \rrbracket_{r, \Delta, \theta} \wedge \Delta(r)(\C{Alive}) = 1\} \quad  \theta' = \theta[v \rightarrow s]\\ \varepsilon' = \varepsilon \cup \{\C{rd}(r,f',n)\ |\ f' \in \mathcal{F}(\phi) \wedge \llbracket f' \rrbracket_{r, \Delta, \theta} = n \} \\ \cup \{\C{rd}(r, f, n)\ |\ \llbracket \phi \rrbracket_{r, \Delta, \theta} \wedge \llbracket f \rrbracket_{r, \Delta, \theta} = n \wedge n \leq s\} \\ \cup \{\C{rd}(r, \C{Alive}, n)\ |\ \llbracket \phi \rrbracket_{r, \Delta, \theta} \wedge \llbracket f \rrbracket_{r, \Delta, \theta} \leq s \wedge \Delta(r)(\C{Alive}) = n\}}{(t, \varepsilon, \C{SELECT} \; \C{MIN}\; f \;\C{AS}\; v\; \C{WHERE}\; \phi, \Delta, \Sigma_r, \theta) \rightarrow (t, \varepsilon', \C{SKIP}, \Delta, \Sigma_r, \theta')}
$$

\rulelabel{E-Sequence}
$$
\RULE{(t, \varepsilon, c1, \Delta, \Sigma_r, \theta) \rightarrow (t, \varepsilon', \C{SKIP}, \Delta, \Sigma_r, \theta')}{(t, \varepsilon, c1;c2, \Delta, \Sigma_r, \theta) \rightarrow (t, \varepsilon', c2, \Delta, \Sigma_r, \theta')}
$$

\rulelabel{E-If-True}
$$
\RULE{\llbracket \phi_c \rrbracket_{\theta} \quad (t, \varepsilon, c1, \Delta, \Sigma_r, \theta) \rightarrow (t, \varepsilon', \C{SKIP}, \Delta, \Sigma_r, \theta')}{(t, \varepsilon, \C{IF}\; \phi_c\; \C{THEN}\; c_1\; \C{ELSE}\; c_2, \Delta, \Sigma_r, \theta) \rightarrow (t, \varepsilon', \C{SKIP}, \Delta, \Sigma_r, \theta')}
$$

\rulelabel{E-If-False}
$$
\RULE{\neg \llbracket \phi_c \rrbracket_{\theta} \quad (t, \varepsilon, c2, \Delta, \Sigma_r, \theta) \rightarrow (t, \varepsilon', \C{SKIP}, \Delta, \Sigma_r, \theta')}{(t, \varepsilon, \C{IF}\; \phi_c\; \C{THEN}\; c_1\; \C{ELSE}\; c_2, \Delta, \Sigma_r, \theta) \rightarrow (t, \varepsilon', \C{SKIP}, \Delta, \Sigma_r, \theta')}
$$

\rulelabel{E-Foreach-1}
$$
\RULE{r \in \theta(v_2) \quad \theta' = \theta[v_1 \rightarrow r][v_2 \rightarrow \theta(v_2) \setminus \{r\}] }{(t, \varepsilon, \tt{FOREACH}\; v_1\; \tt{IN}\; v_2\; \tt{DO}\; c\; \tt{END}, \Delta, \Sigma_r, \theta) \rightarrow (t, \varepsilon', c;\tt{FOREACH}\; v_1\; \tt{IN}\; v_2\; \tt{DO}\; c\; \tt{END}, \Delta, \Sigma_r, \theta')}
$$

\rulelabel{E-Foreach-2}
$$
\RULE{\theta(v_2) = \{\}}{(t, \varepsilon, \tt{FOREACH}\; v_1\; \tt{IN}\; v_2\; \tt{DO}\; c\; \tt{END}, \Delta, \Sigma_r, \theta) \rightarrow (t, \varepsilon', \C{SKIP}, \Delta, \Sigma_r, \theta)}
$$

The definitions of $\M{F}$, $\llbracket \phi \rrbracket_{r, \Delta, \theta}$ (used in the rules of the operational semantics) are presented below:
\begin{mathpar}
\begin{array}{lcl}
\mathcal{F}(\phi_1 \circ \phi_2) & = & \mathcal{F}(\phi_1) \cup \mathcal{F}(\phi_2)\\
\mathcal{F}(\neg \phi) & = & \mathcal{F}(\phi)\\
\mathcal{F}(f \odot e) & = & \{f\} \cup \mathcal{F}(e)\\
\mathcal{F}(e1 \oplus e2) & = &  \mathcal{F}(e1) \cup \mathcal{F}(e2)\\
\mathcal{F}(f) & = & \{f\}\\
\mathcal{F}(v) & = & \phi \\
\mathcal{F}(n) & = & \phi \\
\end{array}
\end{mathpar}

\begin{mathpar}
\begin{array}{lcll}
\llbracket \phi_1 \circ \phi_2 \rrbracket_{r, \Delta, \theta} & = & \llbracket \phi_1 \rrbracket_{r, \Delta, \theta} \circ \llbracket \phi_2 \rrbracket_{r, \Delta, \theta}\\
\llbracket \neg \phi \rrbracket_{r, \Delta, \theta} & = & \neg \llbracket \phi \rrbracket_{r,\Delta, \theta}\\
\llbracket f \odot e \rrbracket_{r, \Delta, \theta} & = & \llbracket f \rrbracket_{r, \Delta, \theta} \odot \llbracket e \rrbracket_{r, \Delta, \theta} \\
\llbracket e_1 \oplus e_2 \rrbracket_{r, \Delta, \theta} & = & \llbracket e_1 \rrbracket_{r, \Delta, \theta} \oplus \llbracket e_2 \rrbracket_{r, \Delta, \theta}\\
\llbracket f \rrbracket_{r, \Delta, \theta} & = &  \Delta(r)(f) & \F{IF}\ f \not\in \C{PK}\\
\llbracket f \rrbracket_{r, \Delta, \theta} & = &  r(f) & \F{IF}\ f \in \C{PK}\\
\llbracket v \rrbracket_{r, \Delta, \theta} & = &  \theta(v)\\
\llbracket n \rrbracket_{r, \Delta, \theta} & = &  n\\
\end{array}
\end{mathpar}

\begin{mathpar}
\begin{array}{lcl}
\llbracket \phi_1 \circ \phi_2 \rrbracket_{\theta} & = & \llbracket \phi_1 \rrbracket_{\theta} \circ \llbracket \phi_2 \rrbracket_{\theta}\\
\llbracket \neg \phi \rrbracket_{\theta} & = & \neg \llbracket \phi \rrbracket_{\theta}\\
\llbracket v \odot e \rrbracket_{\theta} & = & \theta(v) \odot \llbracket e \rrbracket_{\theta}\\
\llbracket v \rrbracket_{\theta} & = &  \theta(v)\\
\llbracket e_1 \oplus e_2 \rrbracket_{\theta} & = & \llbracket e_1 \rrbracket_{\theta} \oplus \llbracket e_2 \rrbracket_{\theta}\\
\llbracket n \rrbracket_{\theta} & = &  n\\
\llbracket \C{NULL} \rrbracket_{\theta} & = &  \{\}\\
\end{array}
\end{mathpar}

\section{Proofs}

\textbf{Lemma 4.} Given an abstract execution $\chi = (\Sigma, \F{vis}, \F{ar})$ and its dependency graph $G_{\chi} = (\Sigma, E)$, the following are true:
\begin{itemize}
\item If $\sigma \xrightarrow{\F{WR}_{r,f}} \sigma' \in E$, then $\sigma \xrightarrow{\F{vis}} \sigma'$.
\item If $\sigma \xrightarrow{\F{WW}_{r,f}} \sigma' \in E$, then $\sigma \xrightarrow{\F{ar}} \sigma'$.
\item If $\sigma \xrightarrow{\F{RW}_{r,f}} \sigma' \in E$, then $\neg (\sigma' \xrightarrow{\F{vis}} \sigma)$.
\end{itemize}

\begin{proof}
The first two statements follow trivially from the definition.
For the third statement, assume for the sake of contradiction that $\sigma' \xrightarrow{\F{vis}} \sigma$. By definition, there exists $\sigma''$ such that $\sigma'' \xrightarrow{\F{WR}_{r,f}} \sigma'$ and $\sigma'' \xrightarrow{\F{WW}_{r,f}} \sigma'$. However, this implies that $\F{MAX}_{\F{ar}}[\F{vis}^{-1}(\sigma)]_{<\F{wri}(r,f)>} = \sigma'$, which contradicts $\sigma'' \xrightarrow{\F{WR}_{r,f}} \sigma'$. Hence, $\neg (\sigma' \xrightarrow{\F{vis}} \sigma)$
\end{proof}
\textbf{Theorem 6.} Given an abstract execution $\chi = (\Sigma, \F{vis}, \F{ar})$, if there is no cycle in the dependency graph $G_{\chi}$, then $\chi$ is serializable.
\begin{proof}
Let $\F{vis'}$ and $\F{ar'}$ be a total order on $\Sigma$ obtained by performing a topological sort of $G_{\chi}$. Now consider the execution $\chi' = (\Sigma, \F{vis'}, \F{ar'})$. It satisfies $\Psi_{\F{Ser}}$. Since $\F{ar'}$ is in the direction of all the $\F{WW}$ edges in $G_{\chi}$, the $\F{WW}$ edges in $G_{\chi'}$ will be the same as $G_{\chi}$. The $\F{WR}$ edges in $G_{\chi'}$ must also be the same as $G_{\chi}$, because otherwise, if $\sigma \xrightarrow{\F{WR}} \sigma' \in G_{\chi'}$ but $\sigma \xrightarrow{\F{WR}} \sigma' \not\in G_{\chi}$, then $\sigma' \xrightarrow{\F{RW}} \sigma \in G_{\chi}$ and hence $\sigma' \xrightarrow{\F{vis'}} \sigma$, which contradicts $\sigma \xrightarrow{\F{WR}} \sigma' \in G_{\chi'}$ (since $\F{vis'}$ is anti-symmetric). Finally, $\F{RW}$ edges in $G_{\chi'}$ will also be the same as $G_{\chi}$, because $\F{WR}$ and $\F{WW}$ edges are the same. Thus, $G_{\chi'}$ and $G_{\chi}$ are isomorphic. Hence $\chi$ is serializable.
\end{proof}
\textbf{Theorem 8.} Given a set of transactional programs $\mathbb{T}$ and a consistency specification $\Psi$, for any valid abstract execution $\chi = (\Sigma, \F{vis}, \F{ar})$ generated by $\mathbb{T}$ under $\Psi$ and its dependency graph $G_{\chi}$, there exists a satisfying model of the formula $\varphi_{\mathbb{T}, \Psi}$ with $\tau = \Sigma$ and the binary predicates $\F{vis}, \F{ar}, \F{WR}, \F{RW}, \F{WW}$ being equal to the corresponding relations in $\chi$ and $G_{\chi}$.
\begin{proof}
Consider a valid abstract execution $\chi = (\Sigma, \F{vis}, \F{ar})$ generated by $\mathbb{T}$ under $\Psi$ and its dependency graph $G_{\chi}$. We start with the model $\M{M}$ where the transaction instances $\tau = \Sigma$, and the binary predicates $\F{vis}, \F{ar}, \F{WR}, \F{RW}, \F{WW}$ are equal to the corresponding relations in $\chi$ and $G_{\chi}$. 

Since the formula $\varphi_{\mathbb{T}, \Psi}$ is a conjunction of clauses, we now prove that $\M{M}$ satisfies all the clauses individually: 

$\varphi_{basic}$ : By the definition of a valid abstract execution, there exists a trace of the transition system $\M{S}_{\mathbb{T}, \Psi}$ of the form $(\{\}, \{\}, \{\}, \{\}) \rightarrow^{*} (\Sigma, \F{vis}, \F{ar}, \{\})$. The \rulelabel{E-Commit} rule is the only rule which allows commited transaction instances to be added to $\Sigma$. It ensures that $\F{ar}$ relation is total, irreflexive, anti-symmetric and transitive, and this can be proved by a simple induction on the length of the trace. Similarly, this rule also ensures that $\F{vis}$ is anti-symmetric and irreflexive. It ensures that $\F{vis}$ implies $\F{ar}$. By Lemma 4, the relations between the dependency edges and $\F{vis}$ and $\F{ar}$ are guaranteed. 

$\varphi_{dep}$ : The rule \rulelabel{E-Spawn} constructs a view of the database obeying the last writer wins condition. Hence, all committed transaction instances in $\Sigma$ will obey this condition. For all dependency relations $\mathcal{R}$, the predicate $\M{R}(r, f, t_1, t_2)$ in the model $\M{M}$ is same as the relation $\M{R}_{r,f}$ in $G_{\chi}$. By the definition of $\M{R}_{r,f}$ (Definition-3), $\varphi_{dep}$ immediately follows.

$\varphi_{\M{R}\rightarrow, \M{T}_1, \M{T}_2}$ : If there exists a dependency edge between two transaction instances $\sigma_1$ and $\sigma_2$, then they must make conflicting database operations to the same field in the same record. Since the database operations result from the execution of SQL statements, it is clear that there must exist SQL statements $\C{c}_1$ and $\C{c}_2$ in $\Gamma(\sigma_1)$ and $\Gamma(\sigma_2)$ which cause the dependency. 

For every type of $\C{c}_1$ and $\C{c}_2$, $\eta_{\C{c}_1, \C{c}_2}^{\M{R}\rightarrow,\M{T}_1, \M{T}_2}$ is a conjunction of the conditionals required for $\C{c}_1$ and $\C{c}_2$ to execute in their respective transaction instances. Since these statements actually execute in $\sigma_1$ and $\sigma_2$, there must exist instantiations for all the variable projection functions (for parameters and local variables) for which these conditionals will be true. If they are inside loops, then there must exist some iteration in which the conflicting operations take place, and hence records in $R$ corresponding to the loop variables. Finally, the rules of the operational semantics (i.e $E-\C{c}_1$ and $E-\C{c}_2$) are designed in such a way that the additional conditions in $\C{c}_2$, $\eta_{\C{c}_1, \C{c}_2}^{\M{R}\rightarrow,\M{T}_1, \M{T}_2}$ depending upon the types of $\C{c}_1$ and $\C{c}_2$ will also be true.

$\varphi_{\rightarrow\M{R}, \M{T}_1, \M{T}_2}$ : Similar reasoning to $\varphi_{\M{R}\rightarrow, \M{T}_1, \M{T}_2}$.

$\varphi_Psi$ : The \rulelabel{E-Commit} rule ensures that the consistency specification is followed at every commit of a transaction instance in $\Sigma$.
\end{proof}
\textbf{Theorem 9.} Given a set of transactional programs $\mathbb{T}$ and a consistency specification $\Psi$, if $\varphi_{\mathbb{T}, \Psi} \wedge \varphi_{D} \wedge \varphi_{\mathrm{Cycle,k}}$ is $\F{UNSAT}$, then all valid abstract executions produced by $\mathbb{T}$ under $\Psi$ of length less than or equal to $\F{k}$ are serializable.
\begin{proof}
By Theorem 8, any valid abstract execution of $\mathbb{T}$ under $\Psi$ is a model of $\varphi_{\mathbb{T}, \Psi}$. Further, any dependency graph which contains a cycle of length less than or equal to $k$ is a model of $\varphi_{D} \wedge \varphi_{\mathrm{Cycle,k}}$. Since $\varphi_{\mathbb{T}, \Psi} \wedge \varphi_{D} \wedge \varphi_{\mathrm{Cycle,k}}$ is $\F{UNSAT}$, this implies that there is no valid abstract execution of $\mathbb{T}$ under $\Psi$ whose dependency graph contains a cycle of length less than or equal to k. By Theorem 6, this implies that all valid abstract executions produced by $\mathbb{T}$ under $\Psi$ of length less than or equal to $k$ are serializable.
\end{proof}
\textbf{Theorem 10.} Given a set of transactional programs $\mathbb{T}$ and a consistency specification $\Psi$, if both $\varphi_{\mathbb{T}, \Psi} \wedge \varphi_{D} \wedge \varphi_{\mathrm{Shortest\ Path,k}}$ and $\varphi_{\mathbb{T}, \Psi} \wedge \varphi_{D} \wedge \varphi_{\mathrm{Cycle,k}}$ are $\F{UNSAT}$, then all valid abstract executions produced by $\mathbb{T}$ under $\Psi$ are serializable. 
\begin{proof}
Since $\varphi_{\mathbb{T}, \Psi} \wedge \varphi_{D} \wedge \varphi_{\mathrm{Cycle,k}}$ is $\F{UNSAT}$, by Theorem 9, all valid abstract execution of $\mathbb{T}$ under $\Psi$ of length less than or equal to $k$ are serializable. Suppose there is a cycle of length greater than $k$ in the dependency graph of a valid abstract execution. Further, assume the cycle is $t \rightarrow \ldots \rightarrow t' \rightarrow t$. Now, the path $t$ and $t'$ must be of length at least $k$. However, since $\varphi_{\mathbb{T}, \Psi} \wedge \varphi_{D} \wedge \varphi_{\mathrm{Shortest\ Path,k}}$ is $\F{UNSAT}$, there does not exist a shortest path of length $k$ in any valid abstract execution. Hence, there must exist a path between $t$ and $t'$ of length strictly less than $k$. This implies a cycle from $t$ to $t$ of length less than or equal to $k$, which contradicts the fact that $\varphi_{\mathbb{T}, \Psi} \wedge \varphi_{D} \wedge \varphi_{\mathrm{Cycle,k}}$ is $\F{UNSAT}$. Hence, there cannot exist a cycle of any length in the dependency graph of a valid abstract execution. By Theorem 6, all valid abstract executions produced by $\mathbb{T}$ under $\Psi$ are serializable.
\end{proof}
\textbf{Lemma 11.} Given a set of transactional programs $\mathbb{T}$, a consistency specification $\Psi$ and a subset of programs $\mathbb{T}' \subseteq \mathbb{T}$, \textbf{if} for all valid executions $\chi$ and their dependency graphs $G_{\chi}$, the following conditions hold:
\begin{enumerate}
\item if $\sigma_1 \xrightarrow{-} \sigma_2$ in $G_{\chi}$ and $\Gamma(\sigma_1) \in \mathbb{T'}$, then $\sigma_1 \xrightarrow{\F{ar}} \sigma_2$
\item if $\sigma_1 \xrightarrow{-} \sigma_2 \xrightarrow{-} \sigma_3$ in $G_{\chi}$, then either $\sigma_1 \xrightarrow{\F{ar}} \sigma_3$ or $\sigma_2 \xrightarrow{\F{ar}} \sigma_3$
\end{enumerate}
\textbf{then} all valid executions which contain at least one instance of a program in $\mathbb{T}'$ are serializable.
\begin{proof}
We first show that for any execution $\chi$ and its dependency graph $G_{\chi}$, if there is a path in $G_{\chi}$ from $\sigma_1$ to $\sigma_2$ such that $\Gamma(\sigma_1) \in \mathbb{T}'$, then $\sigma_1 \xrightarrow{\F{ar}} \sigma_2$. We will show this using strong induction on the length of the path.

\textbf{Base Case}: For paths of length 1, the proof follows trivially from condition (1).

\textbf{Inductive Case}:  Assume that the statement holds for all paths of length less than or equal to $k$. Consider a path of length $k+1$ : $\sigma_1 \rightarrow \sigma_2 \ldots \rightarrow \sigma_{k} \rightarrow \sigma_{k+1} \rightarrow \sigma_{k+2} $. By condition (2), we know that either $\sigma_{k} \xrightarrow{\F{ar}} \sigma_{k+2}$ or $\sigma_{k+1} \xrightarrow{\F{ar}}  \sigma_{k+2}$. By inductive hypothesis, we know that $\sigma_{1} \xrightarrow{\F{ar}} \sigma_{k}$ and $\sigma_{1} \xrightarrow{\F{ar}} \sigma_{k+1}$. By the transitivity of $\F{ar}$, it follows that $\sigma_{1} \xrightarrow{\F{ar}} \sigma_{k+2}$.
Hence, this proves that if there is a path in $G_{\chi}$ from $\sigma_1$ to $\sigma_2$ such that $\Gamma(\sigma_1) \in \mathbb{T}'$, then $\sigma_1 \xrightarrow{\F{ar}} \sigma_2$. This implies that no transaction instance in $\mathbb{T}'$ can be involved in a cycle, because of the irreflexivity of $\F{ar}$. By Theorem 6, this implies that all executions which contain at least one instance of a program in $\mathbb{T}'$ are serializable.
\end{proof}
\textbf{Theorem 12.} Given a set of programs $\mathbb{T}$ and a consistency specification $\Psi$, if $\varphi_{\mathbb{T}, \Psi} \wedge \varphi_{D} \wedge \varphi_{\mathrm{Inductive}, \mathbb{T}'}$ is $\F{UNSAT}$, then all valid executions of $\mathbb{T}$ under $\Psi$ which contains at least one instance of a program in $\mathbb{T}'$ are serializable.
\begin{proof}
If $\varphi_{\mathbb{T}, \Psi} \wedge \varphi_{D} \wedge \varphi_{\mathrm{Inductive}, \mathbb{T}'}$ is $\F{UNSAT}$, then it directly follows that both conditions (1) and (2) of Lemma 11 hold. Hence, all valid executions of $\mathbb{T}$ under $\Psi$ which contains at least one instance of a program in $\mathbb{T}'$ are serializable.
\end{proof}

\end{document}